\documentclass[a4paper,10pt]{article}
\usepackage[position=t,singlelinecheck=off]{subfig}
\usepackage{amssymb,amsmath}
\usepackage{graphics,graphicx,float}
\usepackage{epstopdf}

\usepackage{array}
\newcolumntype{L}[1]{>{\raggedright\let\newline\\\arraybackslash\hspace{0pt}}m{#1}}
\newcolumntype{C}[1]{>{\centering\let\newline\\\arraybackslash\hspace{0pt}}m{#1}}
\newcolumntype{R}[1]{>{\raggedleft\let\newline\\\arraybackslash\hspace{0pt}}m{#1}}

\def\beq{\begin{equation}}
\def\eeq{\end{equation}}
\def\bea{\begin{eqnarray}}
\def\eea{\end{eqnarray}}
\def\nn{\nonumber}

\def\lo{\left(}
\def\rc{\right)}

\def\lg{\left\lgroup}
\def\rg{\right\rgroup}
\makeatletter
 
  \def\@cite#1#2{${\mbox{#1\if@tempswa , #2\fi}}$}
\makeatother

 \date{}
\begin{document}
\begin{center}
{\LARGE \sf{Deconstruction and differentiation of  squeezed kitten states  in a qubit-oscillator system}} \\

\bigskip \bigskip\bigskip M. Balamurugan$^{\dagger}$, R. Chakrabarti$^{\ddagger}$ and B. Virgin Jenisha$^{\dagger}$\\

\begin{small}
\bigskip
\textit{
$^{\dagger}$ Department of Theoretical Physics, 
University of Madras, \\
Maraimalai Campus, Guindy, 
Chennai 600 025, India \\}
 \textit{$^{\ddagger}$Chennai Mathematical Institute, H1 SIPCOT IT Park, \\Siruseri, Kelambakkam 603 103, India}\\
\end{small}
\end{center}
\vfill
\begin{abstract}
\end{abstract}
We study the evolution of the  hybrid entangled  squeezed  states of the qubit-oscillator system in the strong coupling domain. Following the adiabatic approximation we obtain the reduced density matrices of the qubit and the oscillator degrees of freedom. The oscillator reduced density matrix is utilized to calculate the quasiprobability distributions such as the 
Sudarshan-Glauber diagonal $P$-representation, the Wigner $W$-distribution, and the nonnegative Husimi $Q$-function. 
The negativity associated with the $W$-distribution acts as a measure of the  nonclassicality of the state.
The existence of the \textit{multiple} time scales  induced by the interaction introduces certain features in 
the bipartite system. In the strong coupling regime the transient evolution to low entropy configurations reveals 
brief emergence of  \textit{nearly pure} kitten states that may be regarded as superposition of uniformly
separated distinguishable squeezed coherent states. However, the quantum fluctuations with a short time period 
engender bifurcation and subsequent rejoining of these peaks in the phase space.
The abovementioned \textit{doubling} of the number of peaks increases  the entropy to its near maximal value.  Nonetheless, these states characterized by 
high entropy values, are endowed with a large negativity of the $W$-distribution  that points towards their non-Gaussian behavior. This may be ascertained 
by the significantly large Hilbert-Schmidt distance between the oscillator state and an ensemble of 
most general statistical mixture of squeezed Gaussian states possessing  nearly identical second order 
quadrature moments as that of the oscillator.\\ 
\vspace{1cm}

{\bf Keyword:} Hybrid entangled squeezed state, Multiple time scales, Hilbert-Schmidt distance,
Reconstruction nearly pure kitten states, 
Non-Gaussian characteristics.
\newpage

\section{Introduction}
\label{intro}
A two-level system (qubit) that interacts with a radiation field represented by a single oscillator mode
 is one of the important models in quantum optics. 
This model has been studied extensively 
under the rotating wave approximation [\cite{JC1963}] that holds good for the regime characterized by a weak coupling as well as  a small detuning between the qubit and the oscillator frequencies. Recent experiments, however, probe the strong qubit-oscillator coupling
domain, where the rotating wave approximation is not valid. 
Experimental realizations such as a nanoelectromechanical resonator capacitively coupled to a Cooper-pair box driven by microwave currents 
[\cite{ABS2002}, \cite{LaHaye2009}], a quantum semiconductor microcavity
 displaying specific signatures of the ultrastrong coupling regime of the light-matter interaction [\cite{Anappara2009}], a flux-biased quantum circuit that utilizes the large inductance of a Josephson junction to produce an ultrastrong coupling with a coplanar waveguide resonator [\cite{ND2010}] fall in this group. Specifically, the superconducting qubits and circuits are adaptable for a wide range 
 of parameters making them the preferred tools for building quantum simulators [\cite{YN2005}-\cite{GAN2014}]. Additionally, the
hybrid integrated circuits comprising of the atoms, spins, cavity photons, and the superconducting qubits with nanomechanical oscillators 
may facilitate fabrication of interfaces [\cite{ZAYN2013}] in quantum communication network.  
 The Hamiltonian of the strongly coupled qubit-oscillator system incorporates terms that do not preserve the excitation number. To analyze them in the regime where the high oscillator frequency dominates over the low (renormalized) qubit frequency, the authors of [\cite{IGMS2005,AN2010}] have advanced an adiabatic approximation scheme that utilizes the separation of the slow and fast degrees of freedom. This enables the decoupling of the full bipartite Hamiltonian into sectors related to each time scale, and admits approximate evaluation [\cite{IGMS2005}] of its eigenvalues and eigenstates. 

\par

The coupled qubit-oscillator system provides entanglement of the microscopic atomic state and the coherent states of the oscillator, say with opposite phases, that may be
regarded as distinguishable and macroscopic for sufficiently high value of the coherent state amplitude
[\cite{McDonnell2007}]. These analogs of Schr\"{o}dinger cat state are important in understanding the nature of the  decoherence, the quantum-classical transition, and 
the quantum information processing with continuous variables. For instance, they enable non-destructive measurement [{\cite{Guerlin2007}] of the photon number in a field 
stored in a cavity. These states allow  investigations [\cite{Deleglise2008}] on the effects of decoherence by continuously varying the size of the prepared entangled states. Another important proposition is the quantum bus structure [\cite{Loock2008}]  where the hybrid entanglement is used as a tool to eliminate the direct qubit-qubit interactions while admitting a universally coupled continuous mode that mediates among qubits.  Much experimental activity [\cite{ANLF2015}] is geared to observation of such hybrid entangled state. Realization of micro-macro entangled states via a controllable interaction of a single mode microwave cavity field with a superconducting qubit charge, and the subsequent appearance, caused by the measurement of the charge state of the qubit, of superposed macroscopically distinguished field modes have been proposed 
[\cite{LWN2005}]. Recently, optical hybrid entanglement has been observed [\cite{ZVKL2014}] by the superposition of non-Gaussian operations on distinct modes. The hybrid 
entanglement between two remote nodes residing in Hilbert spaces of different dimensionality has been realized [\cite{Morin2014}] using a measurement based procedure. Controllable and deterministic generation of complex superposition of states are also investigated [\cite{LWN2004}-\cite{Vlastakis2013}] using superconducting circuits.  

\par 

On the other  hand, the squeezed states of the harmonic oscillator involves the reduction of the fluctuation in one quadrature variable below the ground state 
uncertainty. This significant property has been used in quantum metrology towards improving the sensitivity of the interferometers 
[\cite{Goda2008}]. Its role in the recent gravitational wave detection via high-power laser interferometers 
[\cite{LIGO2013}], and other possible 
future astronomical observations such as supernova explosions, makes it a key component for the future experiments. Squeezed states are also important in 
continuous variable quantum key distribution protocols [\cite{Lorentz2001}], and it has been proven [\cite{UF2011}] that they can provide an enhancement compared 
to the coherent states. Moreover, they have recently been used as sensitive detectors for photon scattering recoil events at the single photon level 
[\cite{Hempel2013}]. Viewed in this context, the entangled squeezed wave packets in bipartite and multipartite systems hold much promise both for metrology and for 
continuous variable quantum information. Recently such states were experimentally realized [\cite{Lo2015}] where the authors used the internal state-dependent forces to 
create a superposition of distinct squeezed oscillator wavepackets that are entangled with the electronic states of a single trapped ion.

\par

The experimental scenario makes it imperative  to study the evolution of the entangled squeezed wave packets in strongly coupled  hybrid bipartite dynamical systems.
The initial state is taken to be a superposition of distinct squeezed coherent states of the oscillator  that are entangled with the qubit states. Towards understanding 
the dynamics of the quantum state, we utilize various quasi probability distributions on the phase space as preferred tools. For instance, the quasi probability 
$P$-distribution provides a diagonal representation [\cite{S1963}, \cite{G1963}] of the oscillator reduced density matrix in a coherent state basis.  The Wigner 
$W$-distribution [\cite{W1932}] gives a connection between the classical and quantum dynamics as, unlike the classical (true) probability distributions, it may assume 
negative values. Its negativity [\cite{KZ2004}] serves  as an indicator of the nonclassicality of the quantum state. 
 Moreover, investigations [\cite{MPPN2014}]  on the 
nonclassical properties of the qudit cat states as revealed by their $W$-distributions in a $d$-dimensional truncated Fock space has been performed. In the phase space the nonnegativity of the suitably smoothed 
Husimi $Q$-function [\cite{H1940}] facilitates the construction the semiclassical Wehrl entropy [\cite{W1978}]. Also, it
may be employed to provide a semiclassical measure of relative entropy between two suitable $Q$-functions. In the strong coupling regime 
where a quadratic approximation to the effective interaction holds, a long range quasi periodic time dependence in Wehrl entropy  is manifest. This is much parallel to 
the behavior observed earlier [\cite{MTK1990}-\cite{MBWI2001}] in the studies related to nonlinear self-interacting Kerr-like models. It is evident, say, from the 
Wigner $W$-distribution that at the rational submultiples of the long time period kitten-like configurations 
emerge in the phase space. However, multiple time scales present in our model introduces certain  differences  from earlier studies. Owing its origin to the interaction 
 a short time  quantum fluctuation is superposed on the long time quasi periodic behavior. This triggers a bifurcation and rejoining 
of the kitten states on the phase space. Quantum fluctuations in entropy in the short time period points towards recurrence of 
\textit{almost pure} kitten states equi-rotated in the phase space, which are interspersed between near-maximal entropy states that register a doubling of the number of 
kittens. These high entropy states are, however, highly nonclassical as they are endowed with a large negativity of the $W$-distribution.  Employing the 
Hilbert-Schmidt distance [\cite{DMMW2000}] between two quantum states we study this evolutionary aspect in detail. 

\par

The present manuscript is organized as follows:  Starting with the hybrid entangled state of the qubit-oscillator bipartite system the reduced density matrix of the 
oscillator is considered
under the adiabatic approximation in Sec. \ref{OscillatorDenMat}. This density matrix yields the quasi probability distributions such as the Sudarshan-Glauber
 $P$-representation, the Wigner $W$-distribution, and the Husimi $Q$-function that admits a closed form evaluation in the weak coupling regime. The angular distribution 
 in the phase space [\cite{SSW1989}, \cite{ACTS1992}] obtained via the matrix element of the 
 oscillator density operator explicitly provides structure of the kitten states.
For a later comparison between the quantum states the first two moments of the quadrature variables of the oscillator density matrix are obtained. The 
Sec. \ref{KittenState} is devoted to a detailed study of the kitten states emerging due to the presence of multiple time scales during the  evolution in the strong 
coupling domain. We conclude in Sec. \ref{conclude}.

\section{The reduced density matrices and the phase space distributions}
\label{OscillatorDenMat}
We study a coupled qubit-oscillator system with the  Hamiltonian [\cite{IGMS2005}, \cite{AN2010}] that reads 
in natural units $(\hbar=1)$ as follows:
\beq 
 H = -\frac{\Delta}{2} \sigma_x - \frac{\epsilon}{2} \sigma_z +
 \omega a^{\dagger}\, a + \lambda  \sigma_z \,(a^{\dagger} + a),
 \label{Hamiltonian}
\eeq
where the harmonic oscillator with a frequency $\omega$ is described by the raising and lowering operators 
$(a^{\dagger}, \,a| \hat{n} \equiv a^{\dagger} a)$. The spin variables $(\sigma_x, \,\sigma_z)$ characterize the qubit that is furnished with an energy splitting 
$\Delta$ and an external static bias $\epsilon$. Comparison of various types of the charge and the flux qubits depending on the relative strength of the energy parameters $\Delta$ and $\epsilon$ is given in [\cite{YN2005}]. The qubit-oscillator coupling strength is denoted by $\lambda$.  The Fock states 
$\{\hat{n} |n\rangle = n |n\rangle,\,n = 0, 1,\ldots;\;a \,|n\rangle = \sqrt{n}\,|n - 1\rangle, a^{\dagger}\, 
|n\rangle = \sqrt{n + 1}\,|n + 1\rangle\}$ 
provide the basis for the oscillator, whereas the pair of eigenstates  $\sigma_z |\pm 1\rangle = \pm \,|\pm 1\rangle$ comprise the space of the qubit. 
To study the dynamical evolution of the system, we, in the present work, follow the adiabatic approximation 
[\cite{IGMS2005}, \cite{AN2010}] that is based on the separation of the time scales 
governed by a high oscillator frequency and a comparatively low (renormalized)  qubit splitting: $\omega \gg \Delta$. 

\par

Our choice of the initial state of the hybrid bipartite system admits the oscillator in squeezed coherent state while  the entanglement and the interference phase relationship are  activated by the parameter $\mathrm{c} \in \mathbb{C}$:
\beq
|\psi(0)\rangle = \dfrac{1}{\sqrt{1+|\mathrm{c}|^2}}\left(|1 \rangle | \alpha, \xi \rangle 
+ \mathrm{c} \,|-1\rangle |-\alpha, \xi \rangle\right), 
\quad |\alpha\rangle = \mathrm{D}(\alpha)|0 \rangle,\, |\alpha, \xi  \rangle = 
\mathrm{D}(\alpha) \mathrm{S}(\xi)|0 \rangle, 
\label{init_st}
\eeq
where the displacement  and the squeezing operators, respectively, read 
$\mathrm{D}(\alpha)= \exp(\alpha a^{\dagger}-\alpha^*a), \mathrm{S}(\xi)= \exp((\xi^*a^2-\xi a^{\dagger}{^2})/2),\, 
\alpha = \mathrm{Re}(\alpha) + i \,\mathrm{Im}(\alpha) \in \mathbb{C},\, \xi = r \exp(i\,\vartheta),\; r,\vartheta \in \mathbb{R}$.

The number states yield a mode expansion [\cite{GK2005}] of the squeezed coherent state:
\bea
|\alpha, \xi  \rangle = \sum_{n=0}^{\infty} \mathcal{S}_{n}(\alpha, \xi) |n \rangle,
\quad  \mathcal{S}_{n}(\alpha, \xi)=\frac{1}{\sqrt{n!\;\mu}} 
\Big(\frac{\nu}{2\mu}\Big)^{\tfrac{n}{2}}\; \exp\Big(-\frac12\,|\alpha|^2-\frac{\nu}{2\,\mu}\, \alpha^*{}^2 \Big)\, 
 \mathrm{H}_n\Big(\frac{\mu \alpha + \nu \alpha^{*}}{\sqrt{2\,\mu\,\nu}}\Big),
\label{squeezed-mode}
\eea
where the squeezing parameters read: $\mu=\cosh r, \nu= \exp(i \vartheta)\,\sinh r$, and the Hermite polynomials are given by the generating function: 
$\exp (2\,\mathsf{z} \mathsf{t} - \mathsf{t}^{2}) = \sum_{n=0}^{\infty} \tfrac{\mathrm{H}_{n}(\mathsf{z})\, \mathsf{t}^{n}}{n!}$.

\par
 
Employing the adiabatic approximation [\cite{IGMS2005}] the evolution of the initial state (\ref{init_st}) may be constructed  as:
\bea
|\psi(t)\rangle &=& \dfrac{1}{\sqrt{1+|\mathrm{c}|^2}} \exp\Big(-i\frac{\lambda}{\omega}\Big(\mathrm{Im}(\alpha)-\lambda t \Big)\Big) 
\sum_{n=0}^{\infty} \; \exp(-i n \omega t)\; \mathcal{S}_{n}(\alpha_{+}, \xi)  \times \nn \\
& \times & \Big(\mathsf{C}_n(t)\, |1, n_{+}
\rangle + (-1)^n\, \mathsf{D}_n(t) \, |-1, n_{-}\rangle  \Big) ,
\label{state_evolution}
\eea
where the displaced number states $|n_{\pm}\rangle =\mathrm{D}^{\dagger}\left(\pm \frac{\lambda}{\omega}\right) |n\rangle$ provide 
the decomposition basis. In the derivation of (\ref{state_evolution}) the reflection property 
$\mathcal{S}_{n}(-\,\alpha, \xi) = (-1)^{n}\,\mathcal{S}_{n}(\alpha, \xi)$ has been used. The states $|n_{\pm}\rangle$ possess the overlap structure
\beq
  \langle m_-| n_+ \rangle =
 \begin{cases}
 (-1)^{m-n}\;\exp \lo-\frac{x}{2} \rc  x^{\frac{m-n}{2}} \sqrt{n!/m!} \, \, L_n^{m-n}\lo x\rc 
 & \forall \,m \geq n,\\
 \exp \lo-\frac{x}{2} \rc x^{\frac{n-m}{2}} \sqrt{m!/n!} \, \, L_m^{n-m}\lo x\rc & \forall \,m < n,
\end{cases}
\label{overlapp}
\eeq
where the associated Laguerre polynomials read: $L_{n}^{j}(x) = \sum_{k = 0}^{n} 
\,(-1)^{k}\, \binom{n + j}{n - k}\,\frac{x^{k}}{k!}$.
The time-dependent expansion coefficients in (\ref{state_evolution}) are given by
\beq
\mathsf{C}_n(t)=  \cos \chi_n t+ \; i \;  \dfrac{\tilde{\epsilon} - (-1)^n \,\mathrm{c} \, \delta_n}{ \chi_n} \;  \sin \chi_n t, \quad 
\mathsf{D}_n(t)= \mathrm{c} \,\cos  \chi_n t -\; i \;  \dfrac{\mathrm{c} \; \tilde{\epsilon} 
+ (-1)^n \delta_n}{ \chi_n} \; \sin \chi_n t,
\label{o_density_ab}
\eeq
where $\chi_n=\sqrt{\delta_n^2+\tilde{\epsilon}^2},\,
\delta_n = -\tfrac{\widetilde{\Delta}}{2}  L_{n}(x),\, L_{n}(x) \equiv L_{n}^{0}(x),\,x = \left(2\lambda/ \omega\right)^2,\,
\widetilde{\Delta} = \Delta \exp \lo - x/2\rc,\,
\tilde{\epsilon}=\epsilon/2,\, \alpha_{+}=\alpha + \lambda /\omega$.
The state (\ref{state_evolution}) now directly imparts  the  bipartite density matrix: 
$\rho(t) = |\psi(t)\rangle \langle \psi(t)|$. Its partial trace over the qubit (oscillator) degree of freedom produces the reduced density 
matrices of the oscillator (qubit): 
\beq
\rho_{\mathcal{O}} (t) \equiv \hbox {Tr}_{\mathcal Q}\,\rho(t),\quad \varrho_{\mathcal{Q}} (t) \equiv \hbox {Tr}_{\mathcal O}\,\rho(t).
\label{os_qu_density}
\eeq
The explicit construction of the oscillator reduced density matrix for the state (\ref{state_evolution}) reads
\bea
\rho_{\mathcal{O}} (t) &=& \dfrac{1}{1+|\mathrm{c}|^2} \sum_{n,m=0}^{\infty} \mathcal{S}_{n} (\alpha_{+},\xi)\,\mathcal{S}_{m} (\alpha_{+},\xi)^* 
 \Big ( \mathsf{C}_n(t) \mathsf{C}_m(t)^*|n_+ \rangle \langle m_+ |\nn \\
&& + (-1)^{n+m}
\mathsf{D}_n(t) \mathsf{D}_m(t)^* |n_- \rangle \langle m_- | \Big) \exp \big (-i (n - m)\omega t \big).
\label{o_density_matrix}
\eea
The time evolution of the qubit density matrix assumes the form
\beq
\rho_{\mathcal Q}(t) =
\begin{pmatrix}
 \varrho_{1, 1} &  \zeta\\
\zeta^* &\varrho_{-1, -1}
\end{pmatrix},
\label{q_density_matrix}
\eeq
where the matrix elements are structured as
\bea 
\varrho_{1, 1}&=& \frac{1}{1+|\mathrm{c}|^2} \sum_{n=0}^{\infty} |\mathcal{S}_n(\alpha_+,\xi)|^2\, |\mathsf{C}_n(t)|^2, \quad
\varrho_{-1, -1} =\frac{1}{1+|\mathrm{c}|^2}  \sum_{n=0}^{\infty} |\mathcal{S}_n(\alpha_+,\xi)|^2\, |\mathsf{D}_n(t)|^2, \nn \\
\zeta &=& \frac{1}{1+|\mathrm{c}|^2}  \sum_{n,m=0}^{\infty} \!\! (-1)^m 
\mathcal{S}_n(\alpha_+,\xi) \mathcal{S}_m (\alpha_+,\xi)^* \mathsf{C}_n(t)\mathsf{D}_m(t)^* \langle m_- | n_+ \rangle \,
\exp(-i(n-m)\omega t).
\eea
The reduced density matrices (\ref{o_density_matrix}, \ref{q_density_matrix}) satisfy the trace condition: 
$\mathrm{Tr} \varrho_{\mathcal{O}}(t) = 1, \mathrm{Tr} \varrho_{\mathcal{Q}}(t) = 1$. 
The oscillator reduced density matrix (\ref{o_density_matrix}) facilitate the construction of the quasiprobability distributions.
The pair of eigenvalues of the qubit density matrix 
(\ref{q_density_matrix}) read: $\frac{1}{2} \pm \mathfrak{p},\;\mathfrak{p} = \sqrt{\tfrac{1}{4} - |\rho_{\mathcal Q}|}$, where its determinant is given by
$|\rho_{\mathcal Q}| = \varrho_{1, 1} \varrho_{-1, -1}- |\zeta|^2$.  The eigenvalues allow us to compute its von Neumann entropy  
$S(\rho_{\cal Q}) \equiv - \hbox {Tr} (\rho_{\cal Q} \, \log \rho_{\cal Q})$ as
\beq
S_{\cal Q} = - \left(\frac{1}{2} + \mathfrak{p}\right) \log \left(\frac{1}{2} + \mathfrak{p}\right) - 
\left(\frac{1}{2} - \mathfrak{p}\right) \log \left(\frac{1}{2} - \mathfrak{p}\right),
\label{vN_entropy}
\eeq
which measures  the entanglement and the mixedness of the  bipartite system. It is well-known [\cite{AL1970}] that
if  a composite system, comprising of two subsystems, resides in a pure state, the entropies of both subsystems are equal. 
In the present example this yields: $S_{\cal Q} = S_{\cal O} \equiv S$. 

\par

Another quantity that plays a key role in the analysis of the quantum states is Hilbert-Schmidt distance between any two arbitrary quantum density matrices
$\rho_1$ and $\rho_2$  [\cite{DMMW2000}]:
\beq
d_{\mathrm{HS}}(\rho_{1}, \rho_{2}) = \left(\mathrm{Tr}(\rho_1^2) + \mathrm{Tr}(\rho_2^2) -2\; \mathrm{Tr}(\rho_1 \rho_2)\right)^{1/2}.
\label{def-HS}
\eeq
The above definition allows us to evaluate the  distance between the oscillator states (\ref{o_density_matrix}), say, at times $t_1$ and $t_2$ 
via the following norm and the superposition of states:
\bea
\mathrm{Tr} (\rho_{\mathcal O}(t)^2) &=& \frac{1}{(1+|\mathrm{c}|^2)^2} 
 \sum_{n,m=0}^{\infty} |\mathcal{S}_{n}(\alpha_{+}, \xi)|^2 |\mathcal{S}_{m}(\alpha_{+}, \xi)|^2 \quad \quad \times \nn \\
 & &\times  \Big( |\mathsf{C}_n(t)|^2
  |\mathsf{C}_m(t)|^2 + |\mathsf{D}_n(t)|^2
  |\mathsf{D}_m(t)|^2\Big) + 2 |\zeta(t)|^2,\nn\\
 \mathrm{Tr} (\rho_{\mathcal O}(t_1)\,\rho_{\mathcal O}(t_2)) &=& \frac{1}{(1+|\mathrm{c}|^2)^2} \Big(
 \sum_{n,m=0}^{\infty} |\mathcal{S}_{n}(\alpha_{+}, \xi)|^2 |\mathcal{S}_{m}(\alpha_{+}, \xi)|^2 
 \exp(-i \omega (n-m)(t_1-t_2)) \times \nn \\
 & & \times\Big( \mathsf{C}_n(t_1)\mathsf{C}_m(t_1)^* \mathsf{C}_m(t_2) \mathsf{C}_n(t_2)^{*} 
  + \mathsf{D}_n(t_1)\mathsf{D}_m(t_1)^* \mathsf{D}_m(t_2) \mathsf{D}_n(t_2)^{*}  \Big) \Big ) \nn \\
  && +  |\zeta(t_1,t_2)|^2 + |\zeta(t_2,t_1)|^2,
  \label{HS-12}
\eea
where
\beq
\zeta(t_1,t_2) = \frac{1}{1+|\mathrm{c}|^2} \! \sum_{n,m=0}^{\infty} \! \! (-1)^m 
\mathcal{S}_n(\alpha_+,\xi) \mathcal{S}_m (\alpha_+,\xi)^* \mathsf{C}_n(t_1)\mathsf{D}_m(t_2)^* \langle m_- | n_+ \rangle 
\exp(-i \omega(n t_1-m t_2)).
\label{zeta}
\eeq

\subsection{The Sudarshan-Glauber diagonal $P$-representation}
\label{P_evaluation}
Based on the overcompleteness of the coherent states the  Sudarshan-Glauber  $P$-representation [\cite{S1963, G1963}] is well-known to  admit  a diagonal 
construction of the oscillator density matrix in the coherent state basis:
\bea 
\rho_{\mathcal O} = \int P(\beta, \beta^{*})\, |\beta\rangle \langle\beta| \,\mathrm{d}^2 \beta, 
\label{rho-P}
\eea
where the normalizability condition reads $\int P(\beta, \beta^{*})\, \mathrm{d}^2 \beta = 1$. For an arbitrary quantum state the relation (\ref{rho-P})
may be inverted and the diagonal $P$-representation is uniquely expressed [\cite{M1967}] as the following distribution:
\bea
P(\beta, \beta^{*})=\dfrac{\exp(|\beta|^{2})}{\pi^2}\int \langle -\gamma | \rho_{\mathcal O}|\gamma \rangle\, \exp(|\gamma|^2)\,
 \exp(\beta \gamma^* - \beta ^* \gamma) \,\mathrm{d}^2 \gamma.
 \label{P-explicit}
\eea
For  the reduced oscillator density matrix (\ref{o_density_matrix}) the Fourier transform (\ref{P-explicit}) allows explicit construction of the diagonal
$P$-representation as follows:

\bea
P(\beta, \beta^{*}) &=& \dfrac{1}{1+|\mathrm{c}|^2}\sum_{n,m=0}^{\infty} \dfrac{\mathcal{S}_{n}(\alpha_{+},\xi)
S_{m}(\alpha_{+},\xi)^* }{\sqrt{n!m!}}
\Big((-1)^{n+m} \mathsf{C}_{n}(t)\, \mathsf{C}_{m}(t)^*\, \mathcal{K}_{n,m}(\beta_{+})   \nn \\
& &  
+\mathsf{D}_{n}(t)\, \mathsf{D}_{m}(t)^* \, \mathcal{K}_{n,m}(\beta_{-}) \Big)\exp (-i (n-m) \omega t), \; \;
\label{P_explicit}
\eea
where the tensor valued distributions read $\mathcal{K}_{n,m}(z) =  \exp({|z|^2})\,(\tfrac{\partial}{\partial z})^n
(\tfrac{\partial}{\partial z^*})^m\, \delta^{(2)}(z)$, and the dressed phase space variables are given by $\beta_{\pm}=\beta \pm \frac{\lambda}{\omega}$.
The above  $P$-representation is not positive semidefinite, while being highly singular as it incorporates  derivatives of $\delta$ functions.
This is a typical behavior of nonclassical states.
Due to its manifest singular
nature, using the $P$-representation directly towards producing a quantitative measure of  nonclassicality 
is complicated. So other quasiprobability distributions produced via actions of smoothing Gausssian phase space kernels on $P(\beta, \beta^{*})$ are considered 
in this regard. In particular, the nonsingular Wigner $W$-distribution  exhibits negative values  for the present state (\ref{o_density_matrix}), and plays an important role in the study 
of its nonclassicality.
\subsection{The Wigner $W$-distribution}
\label{W_evaluation}
For an arbitrary oscillator density matrix $\rho_{\mathcal O}$ the Wigner quasiprobability distribution is defined [\cite{W1932}]
via the displacement operator as
\beq
W(\beta, \beta^{*})=\frac{1}{\pi^2} \int \mathrm{Tr}[\rho_{\mathcal O} \,\mathrm{D}(\gamma)]\, \exp(\beta \gamma^*-\beta^* \gamma)\, 
\mathrm{d}^2 \gamma, \quad \int W(\beta, \beta^{*})\, \mathrm{d}^2\beta=1.
\label{W_def}
\eeq
The evaluation of the Wigner function using the definition (\ref{W_def}) is, however, not always easy. An equivalent
series representation of the  distribution $W(\beta, \beta^{*})$ in terms of  the diagonal matrix elements in the displaced number states is known
[\cite{CK1993}]:
\beq
W(\beta, \beta^{*})=\dfrac{2}{\pi} \sum^{\infty}_{k=0} (-1)^k \langle \beta,k|\rho_{\mathcal O}| \beta, k \rangle,
\qquad | \beta ,k \rangle = \mathrm{D}(\beta) | k \rangle.
\label{Wigner_series}
\eeq
Substituting the oscillator density matrix (\ref{o_density_matrix}) in the trace relation (\ref{Wigner_series}) we obtain the time evolution of the 
Wigner function for the initial quasi-Bell states:
\bea
W(\beta, \beta^{*})&=& \dfrac{2}{\pi (1+|\mathrm{c}|^2)} \sum_{k=0}^{\infty} (-1)^k \left \lgroup 
 \left | \sum_{n=0}^{\infty} \mathcal{S}_{n}(\alpha_{+},\xi) \, \mathsf{C}_{n}(t)\,
\mathcal{G}_{n, k}(\beta_{+})\, \exp(-i n \omega  t)\right |^2  \right. \nn \\
&& + \left.   \left | \sum_{n=0}^{\infty} (-1)^n \mathcal{S}_{n}(\alpha_{+},\xi)\,
\mathsf{D}_{n}(t)\, 
\mathcal{G}_{n, k}(\beta_{-})\,\exp(-i n \omega t)\right |^2 \right \rgroup.
\label{wigner_inter}
\eea
The tensor components on the phase space 
$\mathcal{G}_{k, \ell}(z)$ are expressed via  the hypergeometric function as
\beq
\mathcal{G}_{k, \ell} (z) =  \exp\Big(\!-\frac{|z|^2}{2}\Big)\, \frac{z^{*}{}^k \,z^{\ell}}{\sqrt{k!\, {\ell}!}}\,{}_2F_0 
\left( \begin{matrix}
- k, -\ell \\
\underline{\phantom{x}}
\end{matrix}; \;-\frac{1}{|z|^2}\right), \quad
{}_2F_0 \left( \begin{matrix} k, \ell\\
\underline{\phantom{x}}           
          \end{matrix};z\right)=\sum_{r=0}^{\infty} \frac{(k)_r (\ell)_r}{r!} z^r,
\eeq
where $(n)_{r} = \prod_{\ell = 0}^{r-1} (n + \ell)$. Proceeding further, the identity [\cite{M1939}] 
\beq
\sum_{k=0}^{\infty} \frac{(-\mathsf{t})^k}{k!} \;{}_2F_0 \left(\begin{matrix} -n, -k\\
                                                \underline{\phantom{x}}     
                                                    \end{matrix};-\frac{1}{\mathsf{t}} \right)
 {}_2F_0 \left(\begin{matrix} -k,-m\\
              \underline{\phantom{x}}  
               \end{matrix};-\frac{1}{\mathsf{t}} \right)
               =2^{n+m} \exp(-\mathsf{t})\;  {}_2F_0 \left(\begin{matrix} -n,-m\\
                                                 \underline{\phantom{x}}        
                                                        \end{matrix}
;-\frac{1}{4\mathsf{t}} \right)
\label{charlier_identity}
\eeq
following from the bilinear kernel of the Charlier polynomials [\cite{M1939}] allows us to recast the Wigner function (\ref{wigner_inter}) in the following form:

\bea
W(\beta, \beta^{*}) &=& \!\! \frac{2}{\pi(1+|\mathrm{c}|)^2} \!\! \sum_{n,m=0}^{\infty} \! \!
  \; \mathcal{S}_{n}(\alpha_{+},\xi)\mathcal{S}_{m}(\alpha_{+},\xi)^*
  \left \lgroup \mathsf{C}_{n}(t) \mathsf{C}_{m}(t)^*\,\mathcal{G}_{n, m} (2\beta_+) \right. 
       \nn \\ 
&& \quad   +  \; (-1)^{n+m} \mathsf{D}_{n}(t) \mathsf{D}_{m}(t)^*\,
\mathcal{G}_{n, m} (2\beta_-)  
   \left. 
 \right \rgroup  \exp(-i (n-m) \omega t). 
\label{wigner}
\eea
The expression (\ref{wigner}) satisfies the normalization condition. We may also explicitly verify that the smoothing of the singular 
$P$-representation (\ref{P_explicit}) via a Gaussian function of variance $1/2$ reproduce the mode sum in (\ref{wigner}):
\beq
W(\beta, \beta^{*})= \frac{2}{\pi}\int  P(\gamma, \gamma^{*})\,\exp(-2|\beta-\gamma|^2)\, \mathrm{d}^2 \gamma.
\label{P-W}
\eeq
The above convolution relation [\cite{C1998}] acts as a consistency check on our derivations. The Hilbert-Schmidt distance (\ref{def-HS}) between two 
arbitrary density matrices $\rho_1$  and $\rho_2$ may be recast [\cite{DMMW2000}] via their corresponding Wigner distributions $W_{1}(\beta,\beta^*)$ and 
$W_{2}(\beta,\beta^*)$ as follows:
\beq
\left(d_{\mathrm{HS}}(\rho_1, \rho_2)\right)^{2} = \pi \int (W_{1}(\beta, \beta^{*})-W_{2}(\beta, \beta^{*}))^2 \;\mathrm{d}^2 \beta.
\label{HS-W}
\eeq

\par

The initial time limit of the  $W$-distribution (\ref{wigner}) consists of two Gaussian peaks as it is the superposition of two well-separated squeezed coherent 
states:
\bea
W(\beta,\beta^*)\vert_{t=0}&=&\dfrac{2}{\pi(1+|\mathrm{c}|^2)} \Big \lgroup \exp(-2|\mu(\alpha-\beta)+\nu (\alpha^*-\beta^*)|^2) \nn \\
&& +|\mathrm{c}|^2 \exp(-2|\mu(\alpha+\beta)+\nu (\alpha^*+\beta^*)|^2)\Big \rgroup,
\label{W-t0}
\eea
where the summation over the Fourier modes in (\ref{wigner}) is realized via the identity [\cite{R1960}]
\beq
\sum_{n=0}^{\infty} \mathrm{H}_{n+k}(\mathsf{z}) \dfrac{\mathsf{t}^n}{n!} = \exp(2\mathsf{z}\mathsf{t}-\mathsf{t}^2) \, 
\mathrm{H}_k(\mathsf{z}-\mathsf{t}).
\label{H-k}
\eeq
The initial value (\ref{W-t0}) of the $W$-distribution maintains  nonnegativity. As time evolves ($t>0$), however, the distribution  (\ref{wigner}) assumes negative values  demonstrating the nonclassical nature of the state.  
For a suitably strong qubit-oscillator coupling a large number of interacting modes set in. The quantum interference between these modes
give rise to negative values of the $W$-distribution in the zone of the phase space intermediate between the positive peaks.  
The volume of the negative sector of the Wigner function on the phase space is considered as 
a  quantitative measure of nonclassicality of the density matrix [\cite{KZ2004}]: 

\beq
\delta_{W} = \int |W(\beta, \beta^{*})| \, \mathrm{d}^{2} \beta - 1.
\label{W-negativity}
\eeq 
In the following analysis we will employ the negativity $\delta_{W}$ as a fruitful way to distinguish between various oscillator states.

 \subsection{Husimi $Q$-function}
 \label{Q_evaluation}
The Husimi $Q$-function [\cite{H1940}] defined as the diagonal expectation value of  the  oscillator density matrix in an arbitrary coherent state
\beq
Q(\beta, \beta^{*})=\dfrac{1}{\pi} \langle \beta| \rho_{\mathcal O}|\beta\rangle  
\label{Q_def}
\eeq
is a positive semi-definite quantity that obeys the normalization condition on the phase space: $\int Q(\beta, \beta^{*}) 
\, \mathrm{d}^{2}\beta = 1$. The $Q$-function may be regarded [\cite{C1998}] as the convolution of the $W$-distribution with a 
Gaussian kernel possessing a variance $1/2$ on the phase space: 
\beq
Q(\beta,\beta^*)= \frac{2}{\pi}\int   W(\gamma,\gamma^*)  \; \exp(-2|\beta-\gamma|^2)\,\mathrm{d}^2 \gamma,
\label{Q_W}
\eeq
which points towards its physical interpretation as  a `coarse-grained' analog of the  $W$-distribution.
Moreover, the process of `coarse-graining' via positive-definite Gaussian kernels of sufficiently broad variance (equal to unity) directly links [\cite{C1998}] 
the singular $P$-representation with the positive semi-definite Husimi $Q$-function:
\beq
Q(\beta,\beta^*)=\frac{1}{\pi}\int P(\gamma, \gamma^*)  \; \exp(-|\beta-\gamma|^2)\,\mathrm{d}^2 \gamma.
\label{Q_P}
\eeq 
For the oscillator density matrix (\ref{o_density_matrix})  the $Q$-function assumes an explicit positive definite form
\bea
Q(\beta, \beta^{*})=\dfrac{1}{\pi(1+|\mathrm{c}|^2)} \Big (  \exp(-|\beta_{+}|^2)\, | \mathcal{X}|^2 + 
\exp(-|\beta_{-}|^2)\, |\mathcal{Y}|^2
 \Big),
\label{Q_factorized}
\eea
where the direct and alternating Fourier series sums respectively read

\beq 
\mathcal{X} = \sum_{n=0}^{\infty} \dfrac{\mathcal{S}_{n}(\alpha_{+},\xi)\beta_{+}^*{}^{n}}{\sqrt{n!}}\,\mathsf{C}_{n}(t)
\,\exp (-i n \omega t),\quad
\mathcal{Y} = \sum_{n=0}^{\infty} (-1)^{n} \dfrac{\mathcal{S}_{n}(\alpha_{+},\xi) \beta_{-}^*{}^{n}}{\sqrt{n!}}\,
\mathsf{D}_{n}(t)\,\exp (-i n \omega t).
\label{XY_def}
\eeq
 The $Q$-function (\ref{Q_factorized}) of the reduced density matrix  (\ref{o_density_matrix}) does not have any zero 
on the phase space except at asymptotically large radial distances. It, however, assumes sufficiently small positive values in the vicinity of 
the negative phase space domains of the $W$-distribution (\ref{wigner}). In the
strong coupling regime $\frac{\lambda}{\omega} \lesssim 0.1$ the positive semi-definite $Q$-function may be expressed in closed form.
Towards this we adopt the procedure [\cite{IGMS2005}] wherein the Laguerre functions are truncated  by retaining only the 
linear parts: $L_n(x)\approx 1-nx+O(x^2)$). The  Fourier sums (\ref{XY_def}) now admit the following approximate closed form expressions:
\beq
 Q_{_\mathrm{linear}}  =\dfrac{1}{\pi(1+|\mathrm{c}|^2)} \Big (  \exp(-|\beta_{+}|^2)\, | \mathcal{X}_{_{\mathrm{linear}}}|^2 + 
\exp(-|\beta_{-}|^2)\, |\mathcal{Y}_{_{\mathrm{linear}}}|^2
 \Big). 
 \eeq
\bea
\mathcal{X}_{_\mathrm{linear}} &=& \dfrac{1}{\sqrt{\mu}} \exp\Big(-\frac12 \widehat{\alpha} \alpha_+^* \Big)  \Big \lgroup (1-\mathsf{A}(\beta_+^*,\omega_+,t))\Phi(\beta_+^*,\omega_+,t) \exp(i \varepsilon t) \nn \\ 
& &+ \,\mathsf{A}(\beta_+^*,\omega_-,t) \Phi(\beta_+^*,\omega_-,t) \exp(-i \varepsilon t) +
\mathrm{c} \; \mathsf{B}(\beta_+^*,\omega_+,t)\Phi(-\beta_+^*,\omega_+,t) \exp(i \varepsilon t)  \nn \\
&  &-\, \mathrm{c} \; \mathsf{B}(\beta_+^*,\omega_-,t)\Phi(-\beta_+^*,\omega_-,t) \exp(-i \varepsilon t)\Big \rgroup,  \nn \\
\mathcal{Y}_{_\mathrm{linear}} &=& \dfrac{1}{\sqrt{\mu}} \exp\Big(-\frac12 \widehat{\alpha} \alpha_+^* \Big)  \Big \lgroup \mathrm{c} \; \mathsf{A}(-\beta_-^*,\omega_+,t) \Phi(-\beta_-^*,\omega_+,t) \exp(i \varepsilon t) \nn \\ 
& &+ \,\mathrm{c} \; (1-\mathsf{A}(-\beta_-^*,\omega_-,t)) \Phi(-\beta_-^*,\omega_-,t) \exp(-i \varepsilon t) +
 \mathsf{B}(-\beta_-^*,\omega_+,t)\Phi(\beta_-^*,\omega_+,t) \exp(i \varepsilon t)  \nn \\
&  &- \,\mathsf{B}(-\beta_-^*,\omega_-,t)\Phi(\beta_+^*,\omega_-,t) \exp(-i \varepsilon t)\Big \rgroup, 
 \label{Q_linear}
\eea
where the coefficients read 
$\varepsilon = \tilde{\epsilon}+\dfrac{{\widetilde{\Delta}}^2}{8\tilde{\epsilon}}, \; 
\omega_{\pm}=\omega \pm \dfrac{{\widetilde{\Delta}}^2}{4\tilde{\epsilon}} x, \; 
\widehat{\alpha}= \alpha_+ + \dfrac{\nu}{\mu}\alpha_+^*,$
and the oscillatory functions assume the form
\bea
\mathsf{A}(\varpi,\mathsf{w},t) &=& \dfrac{{\widetilde{\Delta}}^2}{16\tilde{\epsilon}^2}
-\dfrac{{\widetilde{\Delta}}^2 x \varpi}{8\tilde{\epsilon}^2} \exp (-i\mathsf{w} t) \Big ( \widehat{\alpha}+ \dfrac{\nu}{\mu} \varpi \exp(-i \mathsf{w} t)\Big), \nn \\
\mathsf{B}(\varpi,\mathsf{w},t) &=& \dfrac{{\widetilde{\Delta}}}{4\tilde{\epsilon}}
+ \dfrac{{\widetilde{\Delta}} x \varpi}{4\tilde{\epsilon}} \exp (-i\mathsf{w} t) \Big ( \widehat{\alpha}+ \dfrac{\nu}{\mu} \varpi
\exp(-i \mathsf{w} t)\Big), \nn \\
\Phi(\varpi,\mathsf{w},t) &=& \exp \Big(\widehat{\alpha}\, \varpi \exp(-i \mathsf{w} t)-\dfrac{\nu}{2\mu} 
\varpi^2 \exp(-2i \mathsf{w} t)\Big).
\label{ABPhi}
\eea
\begin{figure}
\begin{center}
\captionsetup[subfigure]{labelfont={sf}}
\subfloat[]{\includegraphics[scale=0.5]{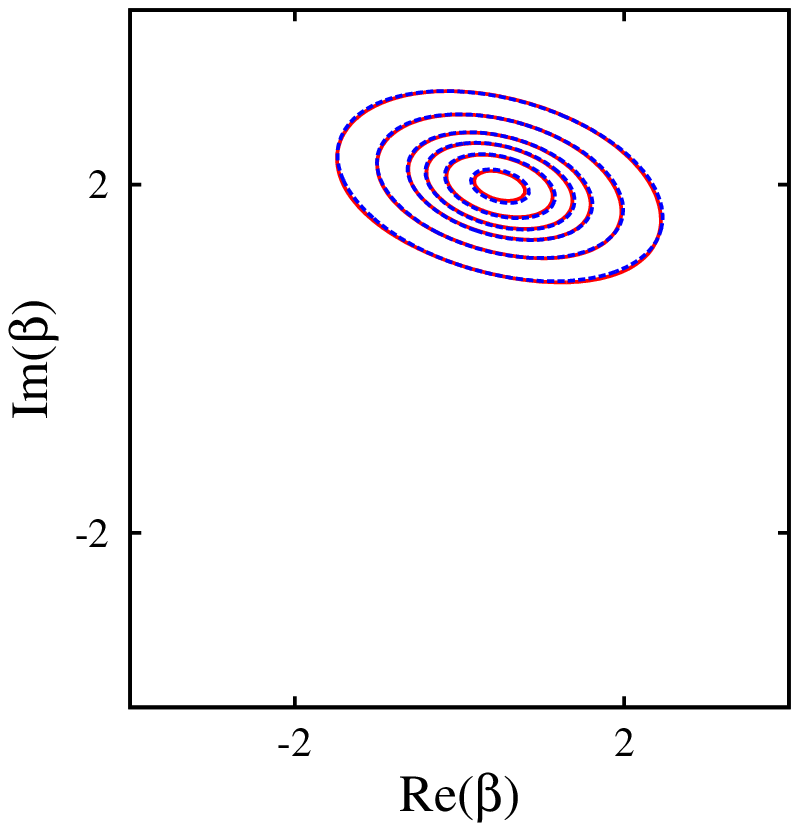}} \quad 
\subfloat[]{\includegraphics[scale=0.5]{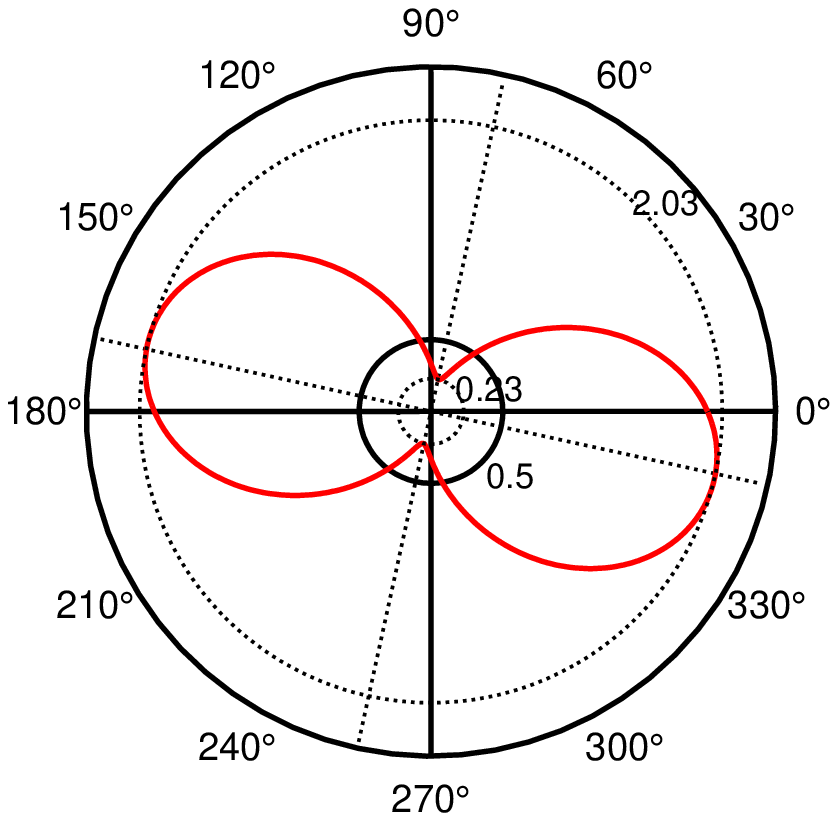}} \quad 
\subfloat[]{\includegraphics[scale=0.5]{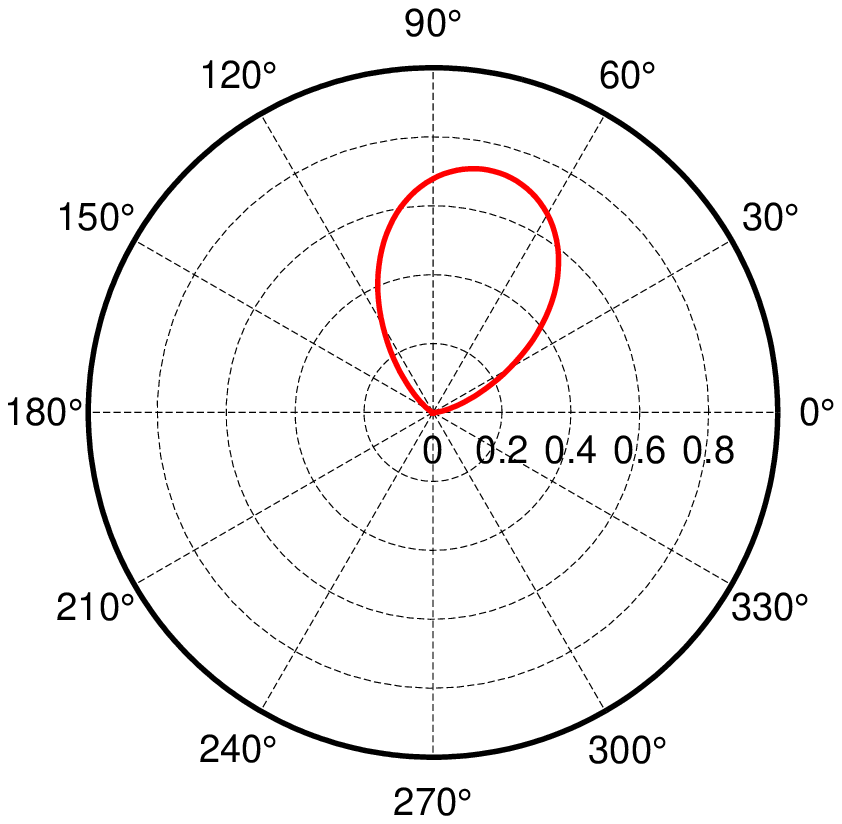}}
\caption{({\sf a}) The $Q$ function for the values $\alpha =2, r=0.7, \vartheta=0,  \Delta =0.15 \; \omega , 
\lambda=0.04 \; \omega, \epsilon= 1.3 \; \omega, \mathrm{c}=1$ at scaled time $\omega t = 30414$ using the (blue dashed) series 
evaluation (\ref{XY_def}), and the corresponding value (red solid) based on the  linear approximation (\ref{Q_linear}).  
 The polar plot of the quadrature variance $V_{\varphi}$ (Sec. \ref{OscillatorDenMat} E) is given in ({\sf b}) for 
the said parameters. ({\sf c}) The angular Husimi density $Q(\theta)$ on the phase space is plotted for  the above set of parameters.}
\label{Q_singlepeak}
\end{center}
\end{figure}
The functional form of $\Phi(\varpi,\mathsf{w},t)$ in (\ref{ABPhi}) suggests that the superposition of Fourier modes leads to  
time-dependent frequencies of oscillation. The deviation between the series evaluation (\ref{XY_def}) and its approximation (\ref{Q_linear}) 
may be measured via the following modulus:
\beq
||\delta Q || = \int |Q-Q_{_{\mathrm{linear}}}| \, \mathrm{d}^2 \beta.
\label{deviation}
\eeq
For the  set of parameters specified in Fig. \ref{Q_singlepeak} its value  $||\delta Q||=0.0214$ specify a $2\%$ error in estimation.

\par

The angular Husimi density $Q(\theta)$ on the phase space [\cite{TMG1993}] is useful in the study of the multiple peaks  characterizing 
kitten-type of states:
\beq
Q(\theta)= \int_{0}^{\infty} Q(\beta,\beta^*) |\beta|\, \mathrm{d} |\beta|, \quad \beta=|\beta|\exp(i\theta).
\label{Q-density}
\eeq
The positive semi-definite angular density is read off the Fourier mode expansion (\ref{Q_factorized}):
\bea
Q(\theta) &=& \!\! \frac{1}{\pi(1+|\mathrm{c}|)^2} \!\! \sum_{n,m=0}^{\infty} \! \!
  \; \mathcal{S}_{n}(\alpha_{+},\xi)\mathcal{S}_{m}(\alpha_{+},\xi)^*
  \left \lgroup \mathsf{C}_{n}(t) \mathsf{C}_{m}(t)^*\,\mathcal{H}_{n, m}^{(+)} (\theta) \right. 
       \nn \\ 
&& \quad   +  \; (-1)^{n+m} \mathsf{D}_{n}(t) \mathsf{D}_{m}(t)^*\,
\mathcal{H}_{n, m}^{(-)} (\theta)
   \left. 
 \right \rgroup  \exp(-i (n-m) \omega t), 
\label{Q_theta}
\eea
where mode sums of the hypergeometric series read:
\bea
\mathcal{H}_{n,m}^{(\pm)} (\theta) &=& \sum_{\jmath=0}^{n}\sum_{\ell=0}^{m} \binom{n}{\jmath} \binom{m}{\ell}
\Big (\pm \dfrac{\lambda}{\omega} \Big)^{n+m-\jmath-\ell} \exp (-i \theta (\jmath-\ell))
\exp \Big (- \dfrac{\lambda^2}{\omega^2} \sin^2 \theta \Big) \times \nn \\
&& \times  \left \lgroup \dfrac{1}{2} \; \Gamma \Big( \dfrac{\jmath+\ell}{2} +1\Big)
{}_1F_1 \left(\begin{matrix} -(\jmath+\ell+1)/2\\
            1/2
               \end{matrix};-\frac{\lambda^2}{\omega^2} \cos^2 \theta \right)\right. \nn \\
& & \mp\, \left. \dfrac{\lambda}{\omega} \cos \theta \; \Gamma \Big( \dfrac{\jmath+\ell+3}{2} \Big)
{}_1F_1 \left(\begin{matrix} -(\jmath+\ell)/2\\
            3/2
               \end{matrix};-\frac{\lambda^2}{\omega^2} \cos^2 \theta \right)
               \right \rgroup.
               \label{Q_H_nm}
\eea
As an example  we plot the angular density $Q(\theta)$ for the observed single-peak in Fig. \ref{Q_singlepeak} ({\sf c}).
\subsection{The angular distribution on the phase space}
A quantity that directly  measures the angular distribution on the phase space corresponding to the oscillator reduced density matrix (\ref{os_qu_density}) is given by [\cite{SSW1989,ACTS1992}]
\beq
\mathcal{P}(\theta)=\frac{1}{2\pi} \langle \theta |\rho_{\mathcal O}| \theta \rangle, \quad |\theta \rangle =\sum_{n=0}^{\infty} \exp(i n \theta)\, |n\rangle,
\label{theta-state}
\eeq
The states $| \theta \rangle$ are nonnormalizable and nonorthogonal for distinct values of the phase angle $\theta$. These states, however,
provide the resolution of unity in the oscillator Hilbert space: $\frac{1}{2 \pi} \int_{0}^{2 \pi} 
\mathrm{d}\theta\, |\theta\rangle\langle \theta| =  \mathbb{I}$. The definition (\ref{theta-state}) via the diagonal element of the density matrix imparts the following properties on the angular distribution function $\mathcal{P}(\theta)$: 
\beq
\mathcal{P}(\theta) \ge 0, \quad \int_{0}^{2 \pi} \mathcal{P}(\theta)\, \mathrm{d}\theta = 1.
\label{P-theta-1}
\eeq
The oscillator reduced density matrix quoted in (\ref{o_density_matrix}) now produces the angular density function as a nonnegative
quantity:
\bea
\mathcal{P}(\theta)=\dfrac{1}{2\pi(1+|\mathrm{c}|^2)} \Big (| \mathsf{X}(\theta,t)|^2 + 
 |\mathsf{Y}(\theta,t)|^2\Big),
\label{Phase_distribution}
\eea
where the Fourier sums may be listed as follows:

\bea 
\mathsf{X}(\theta,t) &=& \sum_{n,k=0}^{\infty}(-1)^k \mathcal{S}_{n}(\alpha_{+},\xi) \,\mathsf{C}_{n}(t) 	\mathcal{G}_{n,k}(\sqrt{x}/2)
\,\exp (-i (n \omega t+k \theta)),\nn \\ 
\mathsf{Y}(\theta,t) &=& \sum_{n,k=0}^{\infty}  \mathcal{S}_{n}(\alpha_{+},\xi) \,
\mathsf{D}_{n}(t) \mathcal{G}_{n,k}(\sqrt{x}/2)\,\exp (-i (n \omega t+k \theta)).
\label{P_XY_def}
\eea
The normalization relation (\ref{P-theta-1}) of the angular density (\ref{Phase_distribution}) may be explicitly checked via the following 
identity [\cite{M1939}] that reflects the orthonormality of the Charlier polynomials:
\beq
\sum_{k=0}^{\infty} \frac{\mathsf{t}^k}{k!} \; {}_2F_0 \left(\begin{matrix} -n, -k\\
                                                \underline{\phantom{x}}     
                                                    \end{matrix};-\frac{1}{\mathsf{t}} \right)
 {}_2F_0 \left(\begin{matrix} -k,-m\\
              \underline{\phantom{x}}  
               \end{matrix};-\frac{1}{\mathsf{t}} \right)
               = n! \;  \mathsf{t}^{-n} \exp(\mathsf{t})\; \delta_{nm}.
\label{charlier_identity_norm}
\eeq
The plots of the angular density distribution $\mathcal{P}(\theta)$ for the kitten states observed at the short time oscillations are given in Fig. \ref{SW_min}.
\subsection{The variance of the quadrature variables}
\label{variance}
While comparing between the phase space distributions associated with distinct density matrices the first and second moments of the quadrature variable 
\beq
X_{\varphi}=\dfrac{1}{\sqrt{2}}(a \exp(-i\varphi) + a^{\dagger} \exp(i\varphi))
\label{EB_def}
\eeq
play an important role. The coordinate and the momentum operators are, respectively, given by 
$\mathsf{q} \equiv \frac{1}{\sqrt{2}}(a + a^{\dagger}) = X_{0}, \mathsf{p} \equiv \frac{1}{\sqrt{2}\,i} (a - a^{\dagger}) 
= X_{\frac{\pi}{2}}$. 
Utilizing our explicit derivation (\ref{Q_factorized}) of the $Q(\beta, \beta^{*})$-function the  expectation values
of the said moments may be obtained: 
\bea
\langle X_{\varphi} \rangle &\equiv& \hbox {Tr}(X_{\varphi} \rho_{\cal O}(t)) = \dfrac{1}{\sqrt{2}} \int 
(\beta \exp(-i\varphi)+ \beta^* \exp(i\varphi)) Q(\beta, \beta^{*}) \mathrm{d}^2 \beta,\nn\\
\langle X_{\varphi}^2 \rangle &\equiv& \hbox {Tr}(X_{\varphi}^2 \rho_{\cal O}(t)) = \dfrac{1}{2} \int
\Big( (\beta \exp(-i\varphi)+ \beta^* \exp(i\varphi))^2-1 \Big ) Q(\beta, \beta^{*}) \mathrm{d}^2 \beta.
\label{moment_def}
\eea
These expectation values read:
\bea
\langle X_{\varphi} \rangle &=& \frac{\sqrt{2}}{1+|\mathrm{c}|^2}\left \lgroup \mathrm{Re} \left ( \sum_{n=0}^{\infty} 
\sqrt{n+1} \; G_{n,1}^{-}(\varphi,t) \right )  - \frac{\sqrt{x}}{2}\,\cos \varphi \, \sum_{n=0}^{\infty} 
G_{n,0}^{-}(\varphi,t)\right \rgroup ,
\label{1st_moment}\\
\langle X_{\varphi}^2 \rangle &=&   \dfrac{1+x \cos^2 \varphi }{2} + 
 |\alpha_+|^{2} + |\nu|^{2} + \frac{1}{1+|\mathrm{c}|^2}
 \mathrm{Re} \left \lgroup   \sum_{n=0}^{\infty} 
\sqrt{(n+1) (n+2)} \; \times \right .  \nn \\
&& \times \left . \; G_{n,2}^{+}(\varphi,t) 
  - 2 \sqrt{x} \,\cos \varphi \, \sum_{n=0}^{\infty} \sqrt{n+1} \; 
G_{n,1}^{+}(\varphi,t)\right \rgroup,
\label{2nd_moment}
\eea
where we employ  the following notations for the coefficients: 
\beq
G_{n,\ell}^{\pm}(\varphi,t) =   \mathcal{S}_n(\alpha_+,\xi) \mathcal{S}_{n+\ell}(\alpha_+,\xi)^* \; 
 \Big ( {\mathsf{C}_{n}}(t) \mathsf{C}_{n+\ell}(t)^* \pm
 \mathsf{D}_{n}(t) {\mathsf{D}_{n+\ell}}(t)^*\Big)\; \exp(i\ell (\omega t+\varphi)).
 \label{G_Fourier} 
\eeq
To compute the moment (\ref{2nd_moment}) the following bilinear kernel of the Hermite polynomials has been used [\cite{R1960}]:
\beq
\sum_{n=0}^{\infty}\dfrac{\mathrm{H}_n(\mathsf{w}) \mathrm{H}_n(\mathsf{z})}{2^n n!} \mathsf{t}^n =  \dfrac{1}{\sqrt{1-\mathsf{t}^2}}
 \exp \Big(\frac{2\mathsf{t}\mathsf{w}\mathsf{z}-\mathsf{t}^2(\mathsf{w}^2+\mathsf{z}^2)}{1-\mathsf{t}^2}\Big). 
\label{H-wz}
\eeq
The variance $V_{\varphi} = \langle X_{\varphi}^2 \rangle - \langle X_{\varphi}\rangle^2$ of the quadrature variable is minimized
at an angle $\varphi_{\mathrm{min}}$ given by $\frac{\partial V_{\varphi}}{\partial \varphi}|_{{\varphi}_{\mathrm{min}}} =0, 
\frac{\partial^{2} V_{\varphi}}{\partial^{2} \varphi}|_{{\varphi}_{\mathrm{min}}} > 0$. Its explicit value reads:
\bea
\varphi_{\mathrm{min}}= \frac12 \Big(  \tan^{-1} \Big( \dfrac{\mathrm{Im}  \langle (\Delta a)^2 \rangle}{\mathrm{Re}  \langle (\Delta a)^2 \rangle} \Big) \pm \pi \Big), 
\quad \langle (\Delta a)^2 \rangle \equiv \hbox {Tr}( \rho_{\cal O}(t)a^2)-(\hbox {Tr}( \rho_{\cal O}(t)a))^2.
\label{phi-Min}
\eea
The state is squeezed along the $X_{\varphi_{\mathrm{min}}}$ quadrature, and stretched along the conjugate direction
 $\varphi_{\mathrm{min}} + \pi/2$ in the 
phase space. The polar plot of the variance $V_{\varphi}$ for a set of parameters
is provided in Fig. \ref{Q_singlepeak} ({\sf b}), where the minimum value of $V_{\varphi}$ reads 0.23,
and the corresponding polar  angle equals $\varphi=76.16^{\circ}$.
We now list the time evolution of the first and the second moments of the quadrature variables [\cite{GP2010}] as follows:
\bea
\langle\mathsf{q}\rangle &=& \langle X_{0} \rangle, \langle\mathsf{p}\rangle = \langle X_{\frac{\pi}{2}} \rangle, 
\sigma_{1 1} \equiv \langle\mathsf{q}^{2}\rangle - \langle\mathsf{q}\rangle^{2}= \langle X^{2}_{0} \rangle -
\langle X_{0} \rangle^{2}, \sigma_{2 2} \equiv \langle\mathsf{p}^{2}\rangle - \langle\mathsf{p}\rangle^{2}= 
\langle X^{2}_{\tfrac{\pi}{2}} \rangle - \langle X_{\frac{\pi}{2}} \rangle^{2},\nn\\
\! \! \sigma_{1 2} \! \!\!\! &\equiv& \!\!\! \!\!\!\dfrac{\langle(\mathsf{q p + p q})\rangle}{2} \! - \! \langle\mathsf{q}\rangle \langle\mathsf{p}\rangle \! \!
= \! \dfrac{1}{1+|\mathrm{c}|^2} \mathrm{Im} \! \! \left \lgroup \! \sum_{n=0}^{\infty} \! \sqrt{n+1} \! \left(\sqrt{x} \; G_{n,1}^{+}(0,t)- \sqrt{n+2} \;
G_{n,2}^{+}(0,t)  \!  \right)\! \! \right \rgroup\!\!\!  - \!\! \langle X_{0} \rangle \langle X_{\frac{\pi}{2}} \rangle. 
\label{M1M2}
\eea
The quadrature moments (\ref{M1M2}) of the time-evolving oscillator states are used in Sec. \ref{KittenState} B towards characterizing their nonclassical properties.

\section{Kitten states in the presence of multiple time scales}
\label{KittenState}
\setcounter{equation}{0}
The evolution (\ref{o_density_matrix}) of the oscillator density matrix $\rho_{\mathcal{O}} (t)$ induces transient appearances of squeezed kitten states 
in the phase space for a strong coupling limit: $\lambda \lessapprox 0.1 \,\omega$. The Wehrl entropy [\cite{W1978}] has been used [\cite{TMG1993}] as an important 
tool for providing the description of these states. Defined as 
\beq
S_{Q} = - \int \, Q(\beta, \beta^{*}) \, \log Q(\beta, \beta^{*})\, \mathrm{d}^{2} \beta,
\label{Wehrl}
\eeq
it measures the delocalization  of the system in the oscillator phase space, and  is considered as a count of an equivalent number of widely separated coherent 
states necessary for tiling the existent occupation on the phase space [\cite{BKK1995}]. As the averaging process via a Gaussian kernel (\ref{Q_W}) plays a key role 
in the construction of the nonnegative $Q$-function, the Wehrl entropy (\ref{Wehrl}) may be envisaged as a quasi classical coarse grained analog [\cite{KMOS2009}] of 
the quantum von Neumann entropy $S$.  In the context of the nonlinear self-interacting Kerr type photonic models, 
the unitary time evolution of a pure coherent state has been found [\cite{MTK1990}-\cite{MBWI2001}] to lead to the formations of the transient kitten states characterized 
by the superposition of a finite number of macroscopic coherent states. In the interaction picture the nonlinearity engenders a periodicity of the Wehrl entropy 
$S_{Q}$ that develops a series of local minima at the rational submultiples of the said time period. Owing to the presence of the interference terms, these 
superposition of multiple coherent states are nonclassical in nature. Recently the  experimental realization [\cite{Kirchmair2013}] of these kitten states has 
been established. The emergence of these transitory kitten states in the \textit{bipartite} qubit-oscillator interacting model studied here has been briefly observed 
earlier [\cite{CJ2015}]. In the current section we provide a detailed investigation of this issue. 
\begin{figure}
\begin{center}
\captionsetup[subfigure]{labelfont={sf}}
 \subfloat[]{\includegraphics[scale=1]{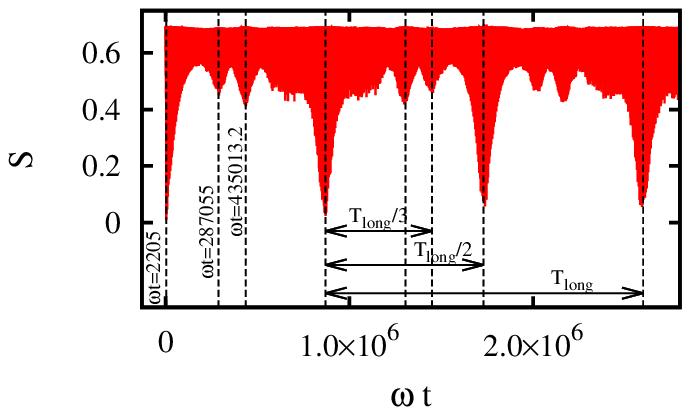}}
 \subfloat[]{\includegraphics[scale=1]{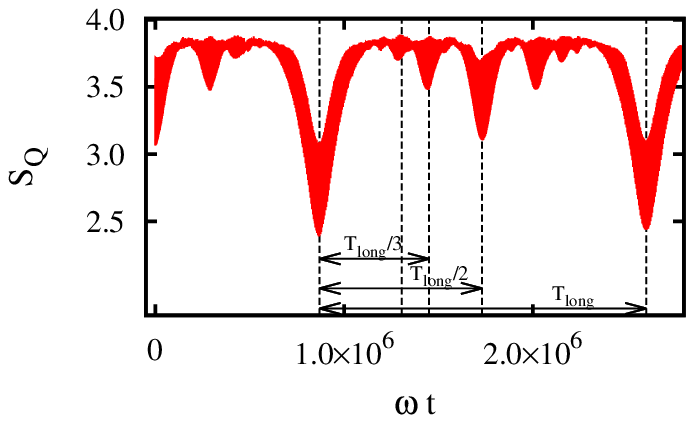}}
 \caption{The long time quasi periodic behavior with the time period $T_{\hbox{\tiny{long}}} \sim O((x^{2} \widetilde{\Delta})^{-1})$ 
 is observed for ({\sf a}) the von Neumann entropy $S$, and ({\sf b}) the  Wehrl entropy $S_Q$ for the parametric 
 values $\alpha =3, r=0.7, 
 \vartheta=0,  \Delta =0.15 \; \omega , \lambda=0.05 \; \omega, \epsilon=0, \mathrm{c}=i$. The semiclassical Wehrl entropy $S_{Q}$ 
 faithfully reproduces periodicity and the local minima structure of the quantum entropy $S$, 
 even though the quantum fluctuations due to a multiplicity of modes originating due to the 
 interaction are more prominent for the latter. The above parameters produce the time period
 $\omega T_{\hbox{\tiny{long}}}$ = 1730000.}
 \label{T-long-S-SQ}
\end{center}
\end{figure}

In the coupling domain $\lambda \lessapprox 0.1 \,\omega$ the interaction frequencies $O(x^{2} \widetilde{\Delta})$ 
and their harmonics are activated giving rise to 
a long time quasi periodic behavior: $T_{\hbox{\tiny{long}}} \sim O((x^{2} \widetilde{\Delta})^{-1})$.
In Fig. \ref{T-long-S-SQ} we observe this quasi periodicity 
in the evolution of \textit{both} the quantum von Neumann entropy $S$, and the quasi classical  Wehrl entropy $S_Q$. The emergence of the long time quasi periodicity, and the occurrence
of the local minima at the rational submultiples of $T_{\hbox{\tiny{long}}}$ are \textit{identically realized} for both 
$S$ (Fig. \ref{T-long-S-SQ} ({\sf a})), and $S_{Q}$ (Fig. \ref{T-long-S-SQ} ({\sf b})). The quantum fluctuations generated by an array of modes are, however, more evident in the 
entropy $S$ rather than in its coarse grained  analog  $S_Q$. Moreover, the existence of particularly low values of the entropy $S$ during its time 
evolution points towards the briefly emerging \textit{almost} pure states of the oscillator subsystem. In contrast to the nonlinear self-interacting Kerr type of models, 
the qubit-oscillator interaction studied here induces a variety of incommensurate modes reflecting a multiplicity of dynamically activated interaction-dependent time 
scales. In the coupling range $\lambda \lessapprox 0.1 \,\omega$, the  modes of frequencies $O(x \widetilde{\Delta})$ cause short range oscillations superimposed on the 
long time quasi periodic behavior observed earlier. In particular, these  linear modes originating via the qubit-oscillator interaction induce an energy transfer between 
the constituent degrees of freedom in a \textit{short} time scale. In the vicinity of the instants marked as the rational submultiples of $T_{\hbox{\tiny{long}}}$ the 
phase space occupation, modulo the fluctuations caused by the modes with frequencies $O(x \widetilde{\Delta})$, achieves a local minimum. The short time period quantum 
fluctuations, however, necessitate spreading on the phase space by splitting the Gaussian peaks (Fig. \ref{SW_min}, columns 2, 4) 
when the energy transfers from the qubit to the 
oscillator mode. The splitting and subsequent rejoining of the Gaussian peaks produced by the interacting modes with frequencies $O(x \widetilde{\Delta})$ cause the local 
fluctuations in the Wehrl entropy $S_Q$ and other dynamical quantities. The splitting of the kittens in the phase space indicates rapidly growing internal complexity of 
the state in the timescale $O((x \widetilde{\Delta})^{-1})$, and is associated with a concomitant growth of  entropies $(S, S_{Q})$. We investigate these issues with 
the choice of the coefficient $\mathrm{c} = i$ in the 
initial hybrid  state (\ref{init_st}) as its time evolution  offers the possibility of creating relatively pure Yurke-Stoler type
[\cite{YS1986}] of squeezed states with sufficiently
high fidelity. As evident in the behavior of the entropies ($S, S_Q$) in Fig. \ref{T-long-S-SQ}, the relative phase $\mathrm{c} = i$ in the  state (\ref{init_st}) 
causes an initial  time translation, compared to the  $\mathrm{c} = \pm 1$ cases, by the amount $T_{\hbox{\tiny{long}}}/2$.
\begin{figure}
\begin{center}
\vspace{-2cm}
\captionsetup[subfigure]{labelfont={sf}}
\subfloat[]{\includegraphics[scale=0.6]{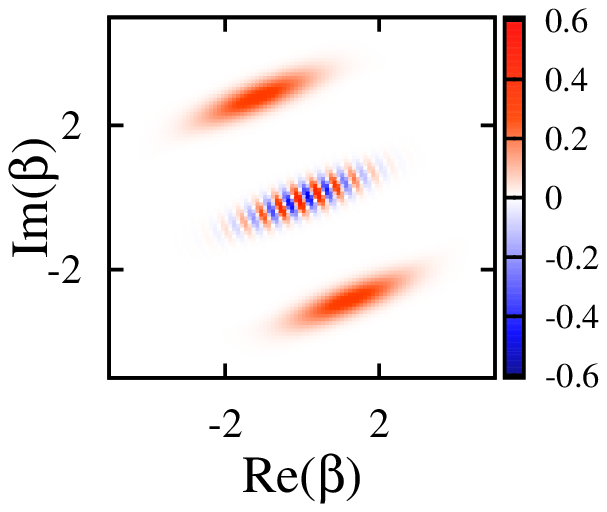}} \hspace{-0.2cm}
\subfloat[]{\includegraphics[scale=0.6]{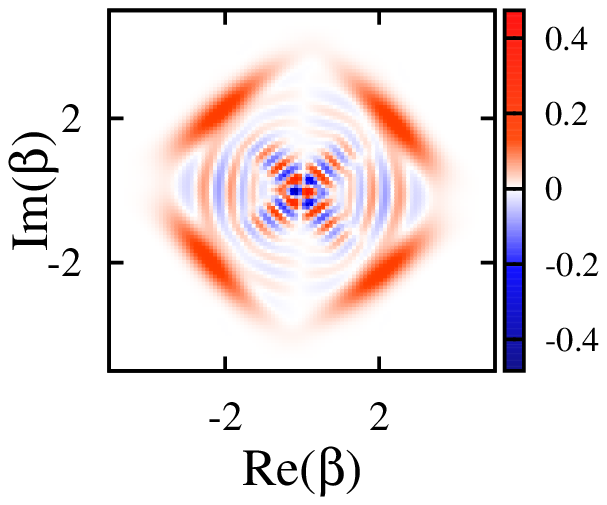}} \hspace{-0.2cm}
\subfloat[]{\includegraphics[scale=0.6]{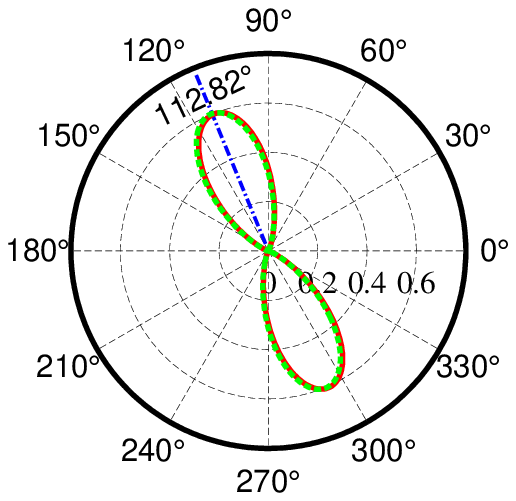}} \hspace{-0.2cm}
\subfloat[]{\includegraphics[scale=0.6]{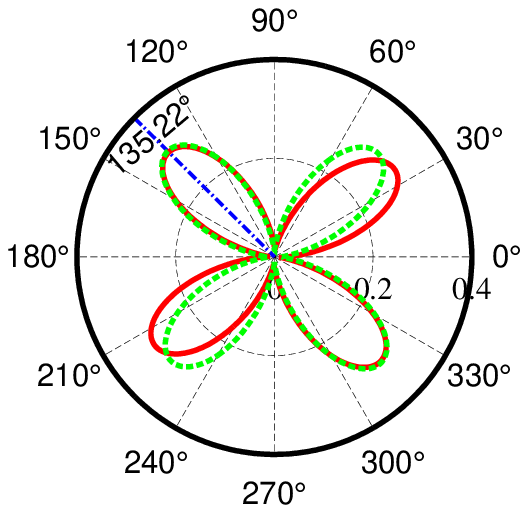}} \hspace{-0.2cm} \\
\subfloat[]{\includegraphics[scale=0.6]{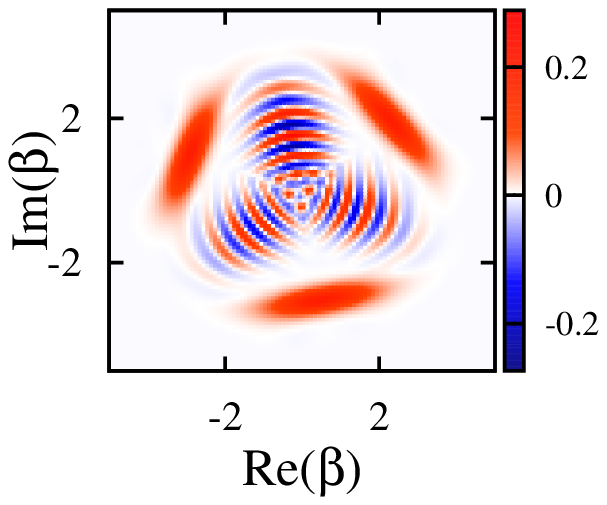}} \hspace{-0.2cm}
\subfloat[]{\includegraphics[scale=0.6]{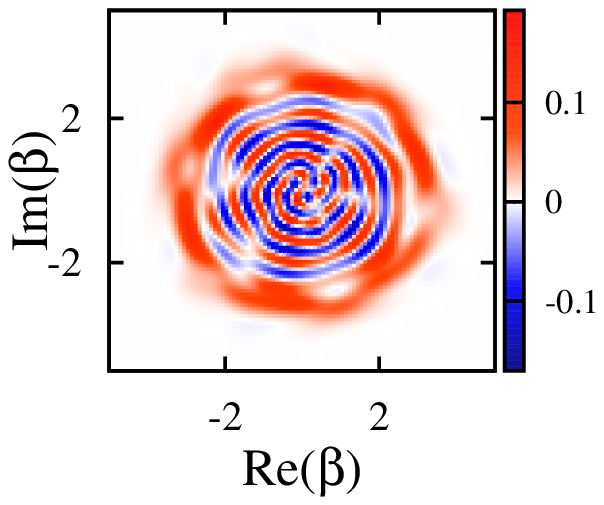}} \hspace{-0.2cm}
\subfloat[]{\includegraphics[scale=0.6]{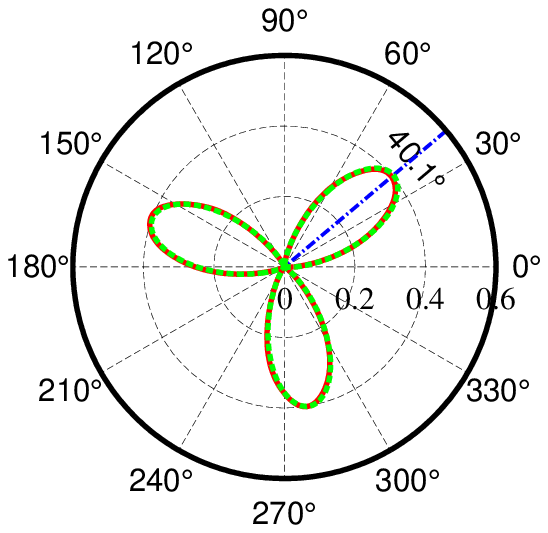}}  \hspace{-0.2cm}
\subfloat[]{\includegraphics[scale=0.6]{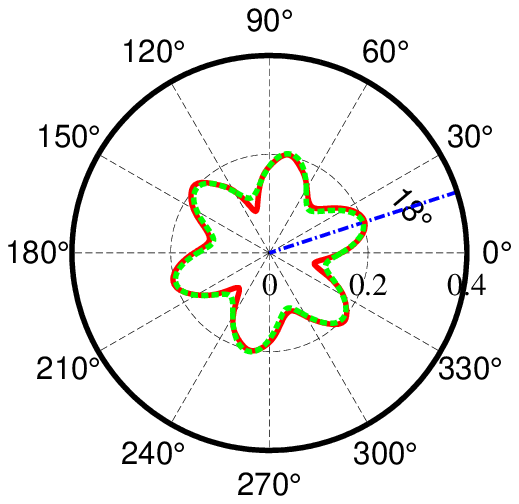}} \\
\subfloat[]{\includegraphics[scale=0.6]{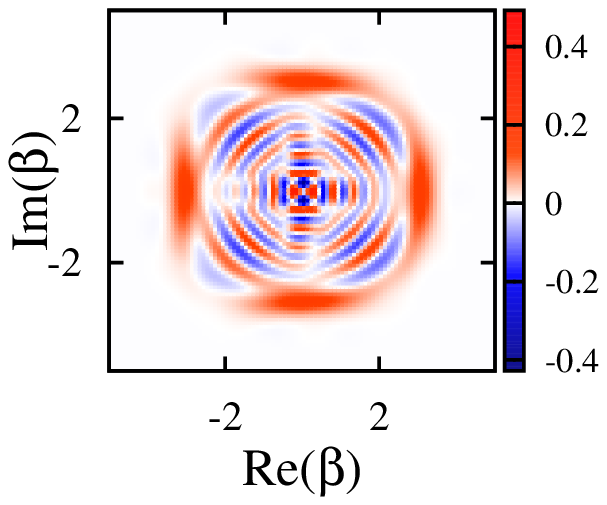}} \hspace{-0.2cm}
\subfloat[]{\includegraphics[scale=0.6]{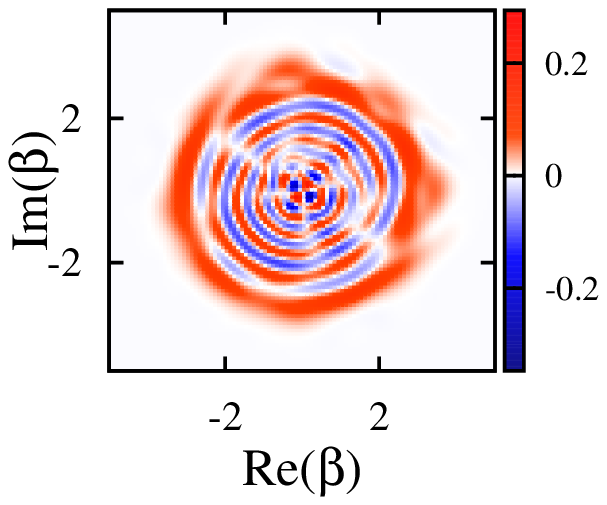}} \hspace{-0.2cm}
\subfloat[]{\includegraphics[scale=0.6]{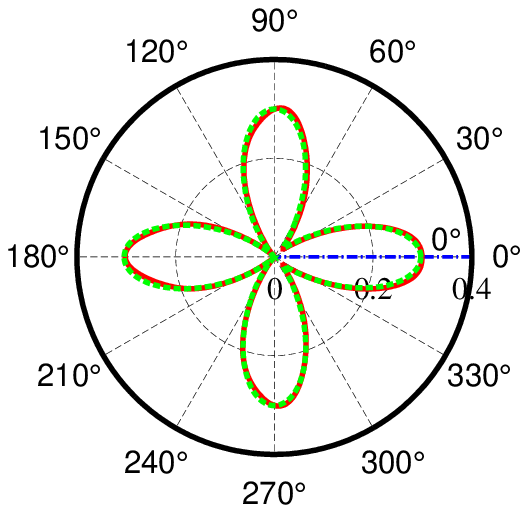}}  \hspace{-0.2cm}
\subfloat[]{\includegraphics[scale=0.6]{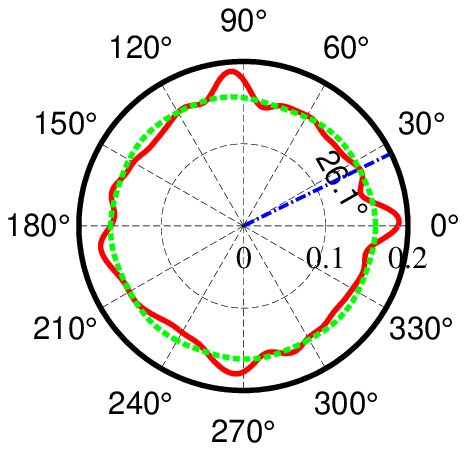}}
\caption{The Wigner $W$-distribution (columns $1, 2$) and the polar plot for
$\mathcal{P}(\theta)$  (columns $2, 4$) at times, equivalent up to a period,  
$T_{\hbox{\tiny{long}}}/2$ (row $1$),  $T_{\hbox{\tiny{long}}}/3$ (row $2$), and  $T_{\hbox{\tiny{long}}}/4$ (row $3$)
respectively. The parameters are chosen as  $\alpha =3, r=0.7, \vartheta=0,  \Delta =0.15 \, \omega, 
\lambda=0.05 \, \omega, \epsilon=0, \mathrm{c}=i$. The short time fluctuations of frequencies
$O(x \widetilde{\Delta})$ superposed on 
the quadratic $T_{\hbox{\tiny{long}}}$ modes
cause the entropies ($S, S_{Q}$) to achieve a local  minimum and a neighboring   maximum at times  $\omega t=2205$ and
$\omega t=3371$ (for  $T_{\hbox{\tiny{long}}}/2$),
 $\omega t=287055$ and $\omega t=286370$
(for  $T_{\hbox{\tiny{long}}}/3$),
 $\omega t=435013.2$ and $\omega t=434450$ (for  $T_{\hbox{\tiny{long}}}/4$), respectively. At the minima of the 
 short time fluctuations at
$T_{\hbox{\tiny{long}}}/p$ give rise to $p$ kitten states that bifurcate into $2 p$ kitten states at the adjacent maxima arrived at the
short time scale $T_{\hbox{\tiny{short}}}$. The green dotted lines in column 3 (4) depict the phase space angular density 
 $\mathcal{P}(\theta)$ of the reference state (\ref{ref-den}) ((\ref{den-th})) with the choice of parameters given in 
 Table \ref{L-Entropy} (\ref{H-Entropy}).}
\label{SW_min}
\end{center}
\end{figure}
In Fig. \ref{SW_min} we  plot the Wigner $W$-distribution and the angular phase space density $\mathcal{P}(\theta)$ at times that, up to a period 
$T_{\hbox{\tiny{long}}}$, are identical with the rational submultiples $T_{\hbox{\tiny{long}}}/2, T_{\hbox{\tiny{long}}}/3, T_{\hbox{\tiny{long}}}/4$, respectively. 
The geometry of the domain on the phase space supporting the $W$-distribution comprises of the Gaussian peaks and their intermediate zones containing the interference 
pattern that exhibits oscillations in a direction perpendicular to the line joining the peaks. As a signature of strong nonclassicality of the state significantly large 
negative domains in the $W$-distribution appear. The short time oscillations of frequency $O(x \widetilde{\Delta})$ are manifest in Fig. \ref{SW_min} as its columns 
$1$ and $3$ ($2$ and $4$) refer, respectively, to the 
$W$-distribution and the polar density plot $\mathcal{P}(\theta)$ to the minimum (maximum) configuration of the quasi classical Wehrl entropy $S_{Q}$. The bifurcation 
 and rejoining of the Gaussian peaks at the time scale 
$T_{\hbox{\tiny{short}}} \sim O((x \widetilde{\Delta})^{-1})$ are clearly evident in these diagrams. Here we make a comment on our  
selection of the coefficient $\mathrm{c} = i$ described above.
Generally speaking, this choice leads to lower entropy states than, say, for alternative values such as $\mathrm{c} = \pm 1$. Moreover, in the former case the odd 
kitten states $(p = 1,3,\ldots)$ are produced for all values of the bias $\epsilon$, whereas in the latter case odd kitten states are realized only for high $\epsilon$. 
This follows from the symmetry 
$Q(\beta, \beta^*)|_{\epsilon = 0} = Q(- \beta, - \beta^*)|_{\epsilon = 0}$ for $\mathrm{c} = \pm 1$ as evident in (\ref{Q_factorized}). One effect of having a high 
bias parameter ($\epsilon \sim \omega$) is that the number of participant interaction-dependent modes will increase much, causing wider fluctuations in the observed 
dynamical quantities.  

We now attempt (Fig. \ref{kittenDouble}) a more detailed  description of the splitting and rejoining of the kitten states in the phase space at time scale 
$T_{\hbox{\tiny{short}}}$. The columns ({\sf a}), ({\sf b}), and ({\sf c}) in  Fig. \ref{kittenDouble} refer, respectively, to the evolution of the 
physical quantities close to the rational submultiples $T_{\hbox{\tiny{long}}}/2, T_{\hbox{\tiny{long}}}/3, T_{\hbox{\tiny{long}}}/4$ of the long period. In the 
rows $1, 2, 3$ and $4$ of  Fig. \ref{kittenDouble} we plot the time evolution at the scale  $T_{\hbox{\tiny{short}}} $ of the Wehrl 
entropy $S_{Q}$, entropy $S$, negativity $\delta_W$, and the Hilbert-Schmidt distance $d_{\mathrm{HS}}$ between quantum states, respectively. In specific, we attempt to reconstruct the states 
where the quantum  entropy $S$ dips to particularly low values signaling evolution to  states  which are \textit{close} to pure states of the oscillator subsystem. 
Moreover, towards establishing the nonclassical nature of  the maximum entropy ($S$) states realized at the time scale
 $T_{\hbox{\tiny{short}}}$, we compare them with the most general statistical mixture of  states possessing nonnegative $W$-distribution, but, nonetheless, 
 endowed with a parallel configuration of Gaussian peaks in the phase space.

\begin{figure}
\begin{center}
\captionsetup[subfigure]{labelfont={sf}}
 \subfloat[]{\includegraphics[scale=0.6]{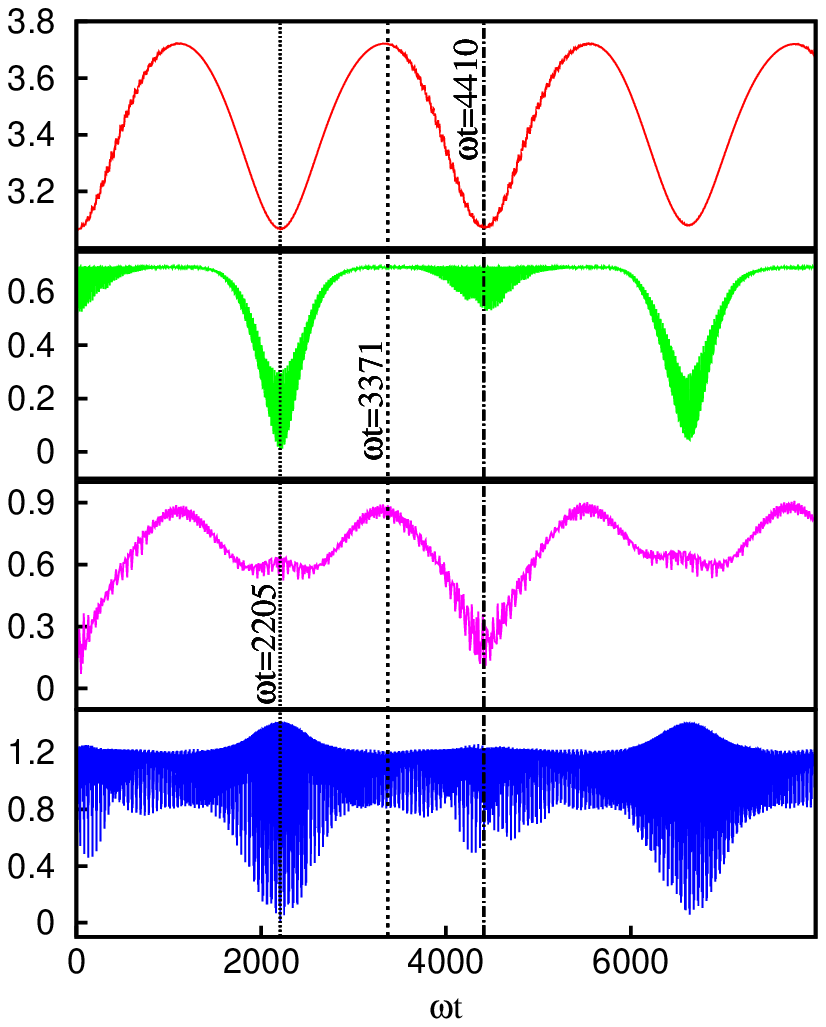}}
\subfloat[]{\includegraphics[scale=0.6]{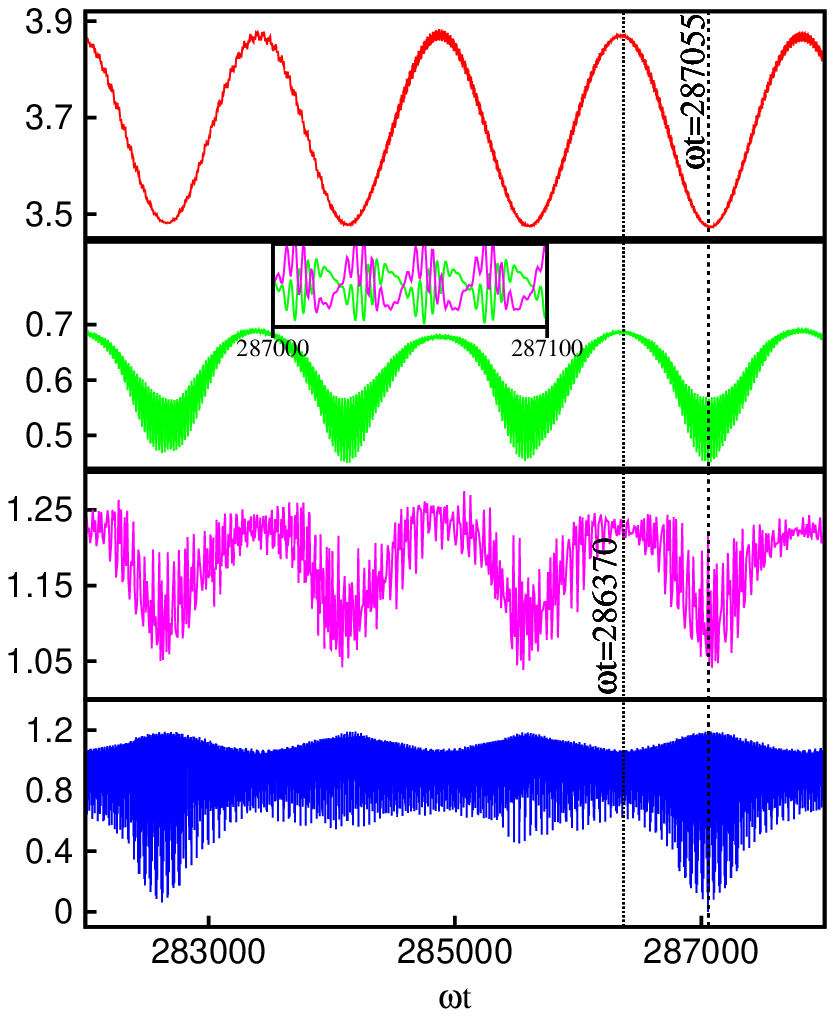}}
\subfloat[]{\includegraphics[scale=0.6]{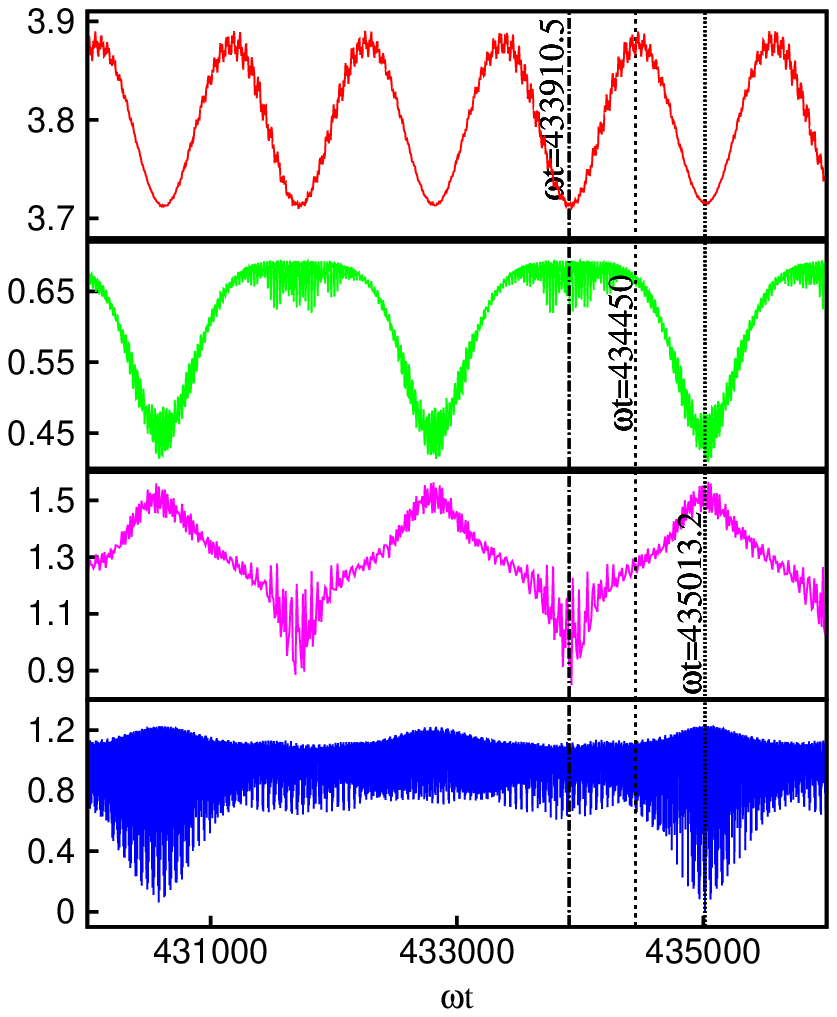}}
\caption{We plot the short time $O(x \widetilde{\Delta})$ fluctuations of the Wehrl entropy $S_Q$(red),  von Neumann entropy 
$S$(green),  negativity $\delta_W$(magenta), and the Hilbert-Schmidt distance 
$d_{\mathrm{HS}}(\rho_{\mathcal O}(t), \rho_{\mathcal O}({t_{{\mathrm{ref}}}}))$(blue) at times 
(equivalent up to a period $T_{\hbox{\tiny{long}}}$) $T_{\hbox{\tiny{long}}}/2$ (column $1$),
$T_{\hbox{\tiny{long}}}/3$ (column $2$), and  $T_{\hbox{\tiny{long}}}/4$ (column $3$) 
respectively. The chosen parametric values read: $\alpha =3, r=0.7, \vartheta=0,  \Delta =0.15 \, \omega, \lambda=0.05 \, \omega, \epsilon=0, \mathrm{c}=i$. 
Towards calibrating the  Hilbert-Schmidt distance $d_{\mathrm{HS}}(\rho_{\mathcal O}(t), 
\rho_{\mathcal O}({t_{{\mathrm{ref}}}}))$ between the quantum states, we, in Figs. ({\sf a}, {\sf b}, {\sf c}),
have chosen the reference times $\omega t_{\mathrm{ref}}$ as  $2205, 287055, 435013.2$, respectively.}
\label{kittenDouble}
\end{center} 
\end{figure}

\subsection{Reconstructing states at the minimum entropy regime}
\label{MinEnt}

The minima realized during the short time oscillations of the Wehrl entropy $S_{Q}$ with frequencies $O(x \widetilde{\Delta})$ at the time limits 
$T_{\hbox{\tiny{long}}}/p,$ where  
$p = 2, 3, \ldots$,  are associated with $p$ kitten states as evident from the corresponding Wigner $W$-distribution (Fig. \ref{SW_min}, column $1$) and the angular phase 
density $\mathcal{P}(\theta)$ (Fig. \ref{SW_min}, column $3$). 
We note that, among the quantities $(S_{Q},S,\delta_W,d_{\mathrm{HS}})$ studied in Fig. \ref{kittenDouble}, the time variation of the  Wehrl entropy $S_{Q}$ 
(Fig. \ref{kittenDouble}, row $1$)  obtained via the smoothed phase space quasiprobability 
$Q$-function is not replicated, in general, by the other variables carrying more quantum informations. For instance, say for the $p = 2, 4$ cases the entropy 
$S$ (Fig. \ref{kittenDouble}, row $2$) possesses domains of its near maximal values even in the regions where the quasi classical quantity $S_{Q}$ is at its  minimum. 
But, interestingly, a \textit{local} minimum in the entropy $S$ \textit {among the adjoining states} develop corresponding to a minimum of its quasi classical analog 
$S_Q$. This feature, where the entropy $S$ undergoes a tiny dip in its value, may be noticed at $\omega t = 4410$ for $p = 2$ (Fig. \ref{kittenDouble} ({\sf a})), 
and $\omega t = 433910.5$ for $p = 4$ (Fig. \ref{kittenDouble} ({\sf c})), 
respectively. A minimum entropy ($S$) configuration indicates relative closeness to a pure state in the Hilbert space, and, in general, 
a consequent decrease in the statistical mixedness of the density matrix.  The increase in the quantum nature of the state  results in a consequent 
enhancement in negativity $\delta_W$ relative to the neighboring states, as observed in Fig. \ref{kittenDouble} ({\sf a}) at time 
$\omega t = 2205$, and at the inset in Fig. \ref{kittenDouble} ({\sf b}) between the time span $\omega t = 287000$ and $\omega t = 287100$.  However, we observe that the 
Hilbert-Schmidt distance $d_{\mathrm{HS}}$ (Fig. \ref{kittenDouble}, row $4$) between the density matrices \textit{most faithfully} distinguishes between the quantum 
states. The  distance $d_{\mathrm{HS}}$ between the quantum states (\ref{def-HS}) are evaluated from the pertinent fiducial  states marked in the 
diagram: $\omega t = 2205$ for $p = 2$ (Fig. \ref{kittenDouble} ({\sf a})), $\omega t = 287055$ for $p = 3$ (Fig. \ref{kittenDouble} 
({\sf b})), and $\omega t = 435013.2$ for 
$p = 4$ (Fig. \ref{kittenDouble} ({\sf c})). In the above cases we notice that the distance  $d_{\mathrm{HS}}$ between the time-evolving oscillator state and the pertinent 
reference marker state approaches 
a null value \textit{only} after a full rotation by the angle $2 \pi$ on the complex plane \textit{i.e.} after completing $p$ cycles 
in the fluctuations of the Wehrl entropy $S_{Q}$ with the period $T_{\hbox{\tiny{short}}}$. 

\par

We now employ the  distance $d_{\mathrm{HS}}$ between the density matrices to determine the proximity of the oscillator  density matrix (\ref{o_density_matrix}) at the 
above low entropy ($S$) limits to the pure states comprising  appropriately rotated and  equi-separated 
$p$ kitten states in the phase space. To facilitate  this reconstruction of a given state we use  an ensemble of density operators that represent a weighted combination 
of ({\sf a}) a dominant pure state  reflecting the said superposition of  $p$  kitten states, as well as ({\sf b}) a comparatively weaker statistical mixture 
of the density matrices of these individual $p$ kitten states. The second term 
 measures the small departure of the state (\ref{o_density_matrix}) from a pure $p$-kitten state density matrix at the chosen times. 
  The construction of the ensemble of the reference states reads:  
\beq
\widetilde{\rho} = \tau \rho_{{}_{\mathrm{pure}}}+ (1-\tau)\rho_{{}_{\mathrm{mixed}}},
\label{ref-den}
\eeq
where the pure state represents a quantum superposition of $p$ equally separated squeezed kitten states: 
\beq
\rho_{{}_{\mathrm{pure}}}= | \psi \rangle \langle \psi |, \quad | \psi \rangle = \dfrac{1}{\sqrt{N_{{}_{\mathrm{pure}}}}}
\sum_{k=0}^{p-1} f_k |\alpha_k, \xi_k\rangle, \; N_{{}_{\mathrm{pure}}}= \sum_{k,\ell=0}^{p-1} f_k f_{\ell}^*  \langle \alpha_{\ell}, \xi_{\ell} | \alpha_k, \xi_k \rangle,
\label{ref-pure}
\eeq
and the mixed state constitutes a statistical mixture of the density matrices of the above squeezed coherent states:
\beq
 \rho_{{}_{\mathrm{mixed}}}=\dfrac{1}{N_{{}_{\mathrm{mixed}}}} \sum_{k=0}^{p-1} g_k|\alpha_k, \xi_k\rangle \langle\alpha_k, \xi_k|, 
\quad N_{{}_{\mathrm{mixed}}}=\sum_{k=0}^{p-1} g_k.
\label{ref-mixed}
\eeq
The phase space coordinates and the squeezing parameters of the $p$ squeezed coherent states employed for the above reconstruction process read:
\beq
\alpha_k = \alpha_+ \exp(i(\widetilde{\vartheta}+2\pi k/p)) \quad
\xi_k = \xi \exp(2 i(\widetilde{\vartheta}+2\pi k/p)), \quad \nu_k = \nu  \exp(2 i(\widetilde{\vartheta}+2\pi k/p)),
\label{alpha-xi-p}
\eeq
where the supplementary angle of rotation $\widetilde{\vartheta}$ has been introduced for implementing maximum phase space overlap between
the density matrix of the reference state (\ref{ref-den}) and that of the transient squeezed kitten states of the density matrix (\ref{o_density_matrix}) observed when the low entropy configurations at $T_{\hbox{\tiny{long}}}/p$ are realized. This is explicitly 
determined from the   angular  distribution function $\mathcal{P}(\theta)$ for the respective cases. The angular distribution 
$\mathcal{P}(\theta)$ of the fiducial state (\ref{ref-den}) are marked in green dotted lines in the Figs. \ref{SW_min} ({\sf c, g, k})
 with the choice of the initial angle $\widetilde{\vartheta}$ and other parameters given in Table \ref{L-Entropy}. In these cases the angular distribution functions $\mathcal{P}(\theta)$ of the two density matrices in comparison make complete overlap. 
 The $\mathcal{P}(\theta)$
 function for the density matrix (\ref{ref-den}) may be easily calculated. We have not explicitly reproduced it here.  The inner product of the squeezed coherent states used in constructing the density matrix (\ref{ref-den}) may be given as below:
\beq
\langle \alpha_{\ell},\xi_{\ell} | \alpha_k, \xi_k \rangle = \dfrac{1}{\sqrt{\mu^2-\nu_k \nu_{\ell}^*}} \exp \Big( -\dfrac{{\mathsf N}_{k \ell}}{\mu^2-\nu_k \nu_{\ell}^*}\Big),
\label{squeeze-inner}
\eeq
where the exponent reads
\beq
{\mathsf N}_{k\ell}= |\alpha_+|^2 (\mu^2+\nu_k \nu_{\ell}^*) -\alpha_k \alpha_{\ell}^*(\mu^2+|\nu|^2) + \dfrac{\mu \nu_k}{2}
 (\alpha_k^*-\alpha_{\ell}^*)^2
+ \dfrac{\mu \nu_{\ell}^*}{2} (\alpha_k-\alpha_{\ell})^2.
\label{squeeze-norm}
\eeq

\par

The equality (\ref{HS-W}) allows us to evaluate the Hilbert-Schmidt distance $d_{\mathrm{HS}}$ between two density matrices via the corresponding Wigner $W$-distributions. Towards this end we now produce the $W$-distribution of the ensemble of the fiducial state (\ref{ref-den}) as follows:
\beq
\widetilde{W}(\beta,\beta^*) =   \tau \; \dfrac{1}{ N_{{}_{\mathrm{pure}}}} \sum_{k, \ell=0}^{p-1} f_k f_{\ell}^* \;  
\widetilde{W}_{k \ell}(\beta,\beta^*) + (1-\tau) \; \dfrac{1}{ N_{{}_{\mathrm{mixed}}}} \sum_{k=0}^{p-1} g_k \; \widetilde{W}_{kk}(\beta,\beta^*),
\label{W-ref}
\eeq
where the component $\widetilde{W}_{k \ell}(\beta,\beta^*)$ associated with the projection operator $|\alpha_k,\xi_k \rangle \langle \alpha_l, \xi_l |$
may be determined \textit{\`{a} la} the series sum given in (\ref{Wigner_series}):
\beq
\widetilde{W}_{k \ell}(\beta, \beta^{*})=\dfrac{2}{\pi} \sum^{\infty}_{n=0} (-1)^n \langle \beta,n|\alpha_k,\xi_k \rangle \langle \alpha_{\ell}, \xi_{\ell} |\beta, n \rangle =  \dfrac{2}{\pi\sqrt{\mu^2-\nu_k \nu_{\ell}^*}}\; 
\exp \Big(- \dfrac{{\mathsf N}_{k \ell} + 2 {\mathsf G}_{k \ell}}{\mu^2-\nu_k \nu_{\ell}^*}\Big),
\label{Wigner-kl}
\eeq
where the Gaussian exponent is given by
\beq
{\mathsf G}_{k \ell} =   \mu^2 (\alpha_k -\beta)(\alpha_{\ell}^* -\beta^*) +  \nu_k \nu_{\ell}^* 
(\alpha_k^* -\beta^*)(\alpha_{\ell} -\beta) 
 + \,\mu \nu_k (\alpha_k^* -\beta^*)(\alpha_{\ell}^* -\beta^*) + \mu \nu_{\ell}^* (\alpha_k -\beta^)(\alpha_{\ell} -\beta).
 \label{G-kl}
\eeq

\par

Following the above procedure we now attempt an explicit reconstruction of the density matrices associated with the transient kitten states as the candidates located in the Hilbert space in  close neighborhood of the corresponding pure states (\ref{ref-pure}). To illustrate the process we choose the following examples at times ({\sf i}) $\omega t=2205$ in Fig. \ref{kittenDouble} 
({\sf a}), ({\sf ii}) $\omega t=287055$ (Fig. \ref{kittenDouble} ({\sf b})), and  ({\sf iii}) $\omega t = 435013.2$ 
(Fig. \ref{kittenDouble} ({\sf c})), where the entropy  $S$ assumes a local minimum value  regarding the fluctuations in the time scale $T_{\hbox{\tiny{short}}}$.  
The parameters defining the fiducial marker state  (\ref{ref-den}) $\{f_k, g_k| k= 0, 1, \ldots, p-1; \tau\}$ 
are varied \textit{independently} to obtain the minimum Hilbert-Schmidt distance between the density matrices $\rho_{\mathcal O}$ and $\widetilde{\rho}$. Even though the 
decomposition process (\ref{ref-den}) and the subsequent reconstruction of the density matrix are not unique, but the variational computation based on minimization of 
the Hilbert-Schmidt distance provides a robust selection process in a large ensemble of states. The results
are summarized in Table \ref{L-Entropy}. We express our results in terms of a dimensionless quantity 
maintaining values $d_{\mathrm{HS}}(\rho_{\mathcal O}(t), \widetilde{\rho})/\sqrt{\mathrm{Tr}(\rho_{\mathcal O}(t)^2)} \ll 1$ at times  when the low entropy states are 
realized. The margin of error in the above analysis is limited by the requirement $\Delta d_{\mathrm{HS}}(\rho_{\mathcal O}(t), \widetilde{\rho}) \le 
|d_{\mathrm{HS}}(\rho_{\mathcal O}(t), \widetilde{\rho})|$.  The data presented in Table \ref{L-Entropy} suggest that as the number of kittens increases the maximum possible 
error is also increased. These examples, considered at the instants corresponding to locally minimum entropy  $S$ configurations, indicate that, modulo a small component of 
statistical mixed state, the oscillator density matrix (\ref{o_density_matrix}) evolves to a generalized Yurke-Stoler [\cite{YS1986}] type of pure state.

\begin{table}[]
\centering
\renewcommand{\arraystretch}{1.2}
\begin{tabular}{|C{1.3cm}|C{0.4cm}|C{1cm}|C{1.2cm}|C{3.1cm}|C{3.1cm}|C{0.9cm}|C{1.1cm}|C{1.2cm}|}
\hline
$\omega t$ & $p$ &$S$   & $\widetilde{\vartheta}$  & $f_k|k=0,1,\ldots, p-1$ & $g_k|k=0,1,\ldots, p-1 $   & $\tau$ & \rotatebox{90}{$d_{\mathrm{HS}}(\rho_{\mathcal O}(t),\widetilde{\rho})\;\;$} &
\rotatebox{90}{$\dfrac{d_{\mathrm{HS}}(\rho_{\mathcal O}(t),\widetilde{\rho})}{\sqrt{\mathrm{Tr}(\rho_{\mathcal O}(t)^2}}\;\;$}    \\ \hline
$2205$   & 2    & 0.0118 & $112.82^{\circ}$ & $ 1$,   $ \exp(i \pi 1.677)$ & $  1, 1 $ & 0.996  & 0.0647   & 0.0649 \\ \hline
 $\!\!\!\!\!287055$      &   3&  0.4495  &  $40.1^{\circ}$ &$ 1$, $0.97 \exp(i \pi 1.315)$, $0.625 \exp(i \pi 1.415) $&       $0, 0.095, 1.275 $ &   0.783     &     
 0.0879     & 0.1034        \\ \hline
   $\!\!\!\!\! 434013.2$ & 4 &  $0.4085$ &  $0^{\circ}$   &    $ 1$, $0.96 \exp (i \pi 0.11)$, $\exp (i \pi 1.005)$, $0.96 \exp(i \pi 1.895)$& $1, 1.54, 1.14, 1.50$ & 0.810       &   0.1467      
    & 0.1685        \\ \hline
    
\end{tabular}
\caption{Reconstruction of oscillator states in the neighborhood of pure states}
\label{L-Entropy}
\end{table}

\subsection{Characterization of states at the large entropy regime}
\label{LrgEn}
In  Fig. \ref{kittenDouble} we observe that the quantum fluctuations of the  frequencies $O(x \widetilde{\Delta})$ give rise to a 
periodic $(T_{\hbox{\tiny short}})$ maximization of the entropy $S$ and the Wehrl entropy $S_Q$.
At the instants when the entropies  are maximized,
the $T_{\hbox{\tiny short}}$ fluctuations occurring at the long range time scales 
$T_{\hbox{\tiny long}}/p$ (up to a period $T_{\hbox{\tiny long}}$) produce, on account of the splitting in the phase space 
distribution,
a combination of $2 p\,(p =  2, 3,4, \ldots)$ squeezed kitten configurations.
In particular, we examine the  oscillator 
phase space distributions at times $\omega t = 3371$ (Fig. \ref{kittenDouble} ({\sf a})), $\omega t = 286370$ 
(Fig. \ref{kittenDouble} ({\sf b})), and
$\omega t = 434450$ (Fig. \ref{kittenDouble} ({\sf c})), which, as it is evident in the Fig. \ref{SW_min} (columns $2, 4$), correspond to $4, 6$, and $8$ squeezed kitten 
configurations, respectively, in the phase space. As  these states possess large entropy ($S$), they are 
necessarily far away from the pure states in the Hilbert space. 
However, they are also endowed with large negativity $\delta_{W}$ (Fig. \ref{kittenDouble}) of the 
Wigner $W$-distribution, and, consequently, they are 
strongly nonclassical in nature.  

\par 

To study the non-Gaussian properties of the above states we 
compare them with the most general statistical mixture of the Gaussian squeezed states, 
where the ensemble exhibits an identical number of kitten configurations in the phase space.
The density matrix of such a state with an array of equally separated $2 p$ kitten combinations 
in the phase space is given by [\cite{GP2010}]
\beq
\rho_{_{\mathrm{mixed}}}^{\hbox{\tiny th}} = \dfrac{1}{N_{\mathrm{mixed}}} \sum_{k=0}^{2 p-1}  g_k\,\mathrm{D}(\alpha_k) \mathrm{S}(\xi_k)  
\rho_{_{\mathrm{th}}}\mathrm{S}^{\dagger}(\xi_k) \mathrm{D}^{\dagger}(\alpha_k),\quad \rho_{_{\mathrm{th}}} =\dfrac{1}{\bar{n} + 1}\,
\left(\dfrac{\bar{n}}{\bar{n} + 1}\right)^{a^{\dagger} a}, \;\;\bar{n} = \dfrac{1}{\exp(\beta_{B} \omega) - 1},
\label{den-th}
\eeq
where $\rho_{_{\mathrm{th}}}$ is equilibrium thermal density matrix, and 
$\beta_{\mathrm{B}}$ is the inverse temperature. 
It has been proposed [\cite{JKL2015}] that the zero temperature limit of the state (\ref{den-th}) may be 
employed towards enhancing the upper bound 
on the accessible information in a Gaussian private quantum channel.
The Wigner $W$-distribution of the Gaussian state (\ref{den-th})
 is nonnegative, and, therefore, it may be used as a suitable benchmark for studying the quantum features 
 of the oscillator density 
 matrix evolving in time.
For later use we quote the  quasi probability distributions of the statistical mixed state (\ref{den-th}) below:
\bea
W(\beta,\beta^*)|_{_{\mathrm{mixed}}} &=&  \dfrac{1}{\pi (\bar{n}+1/2)\,N_{_{\mathrm{mixed}}}}  \sum_{k=0}^{2p-1}\; g_k 
 \exp \left\lgroup- \dfrac{1}{\bar{n}+1/2}|\mu(\alpha_k-\beta)+\nu_k (\alpha_k^*-\beta^*)|^2 \right\rgroup, 
 \label{W_th_mix}\\
Q(\beta,\beta^*)|_{_{\mathrm{mixed}}} &=& \dfrac{1}{\pi\, \mathfrak{N}\,N_{_{\mathrm{mixed}}}} \sum_{k=0}^{2p-1}\; 
g_k \exp \lg -\dfrac{1}{\mathfrak{N}^{2}} 
\Big((\mu^2(1+\bar{n})+|\nu|^2 \bar{n})|\beta-\alpha_k|^2 \right.\nn \\
& &+ \left.\dfrac{\mu}{2}(2\bar{n}+1) (\nu_k^* (\beta-\alpha_k)^2 + \nu_k (\beta^*-\alpha_k^*)^2 )\Big)\rg, \quad 
\mathfrak{N} = \sqrt{\bar{n}^2+\mu^2(2\bar{n}+1)}.
\label{Q_th_mix}
\eea

\par

Towards examining the above mentioned  high entropy states displaying nonclassicality 
\textit {vis a vis} their Gaussian partners now we compute the first two moments of the 
quadrature variables with respect to the statistical mixture given in (\ref{den-th}). 
The underlying idea is to distinguish between the  non-Gaussian high entropy squeezed kitten state configurations, 
and the corresponding Gaussian states endowed with almost identical gross characteristics like the 
first two moments of the quadrature variables [\cite{GP2010}]. The expectation value of a dynamical observable
for the Gaussian statistical mixture (\ref{den-th}) reads: 
$\langle \mathcal{X}\rangle _{_{\mathrm{mixed}}} = \mathrm{Tr} (\mathcal{X} \rho_{_{\mathrm{mixed}}}^{\hbox{\tiny th}})$.  As the moments up to the second order  completely describe the Gaussian states, we quote them below:
\bea
\langle\mathsf{q}\rangle_{_{\mathrm{mixed}}} &=& \dfrac{\sqrt{2}}{N_{_{\mathrm{mixed}}}} \sum_{k=0}^{2 p-1}\; g_k \; 
\mathrm{Re} (\alpha_k ),  \qquad 
  \langle{\mathsf p}\rangle_{_{\mathrm{mixed}}} = \dfrac{\sqrt{2}}{ N_{_{\mathrm{mixed}}}} \sum_{k=0}^{2 p-1} \; g_k \; \mathrm{Im} (\alpha_k ), \nn \\
\langle\mathsf{q}^2 \rangle_{_{\mathrm{mixed}}} &=& \dfrac{1}{N_{_{\mathrm{mixed}}}} \sum_{k=0}^{2 p-1}\; g_k 
\left ((\bar{n}+\tfrac{1}{2}) (\mu^2+|\nu|^2-2\mu\,\mathrm{Re}(\nu_k)) + 2 \left (\mathrm{Re} (\alpha_k )\right )^2  \right),  \nn \\
\langle\mathsf{p}^2 \rangle_{_{\mathrm{mixed}}} &=& \dfrac{1}{N_{_{\mathrm{mixed}}}} \sum_{k=0}^{2 p-1}\; g_k 
\left ((\bar{n}+\tfrac{1}{2}) (\mu^2+|\nu|^2+2\mu\,\mathrm{Re}(\nu_k)) + 2 \left(\mathrm{Im} (\alpha_k ) \right )^2  \right),\nn\\
\tfrac{1}{2}\,\langle(\mathsf{p\,q+q\,p})\rangle_{_{\mathrm{mixed}}} &=& \dfrac{1}{N_{_{\mathrm{mixed}}}} \sum_{k=0}^{2 p-1}\; g_k  \; \mathrm{Im} \Big (\alpha_k^2 - (2\bar{n}+1)  \mu\, \nu_k\Big).
\label{moment-th}
\eea
The second order covariance matrix elements [\cite{GP2010}] for the state (\ref{den-th}) are given by 
\bea
\sigma_{11}|_{_{\mathrm{mixed}}} &=& \langle\mathsf{q}^2 \rangle_{_{\mathrm{mixed}}} - 
\langle\mathsf{q} \rangle^{2}_{_{\mathrm{mixed}}},\quad\sigma_{22}|_{_{\mathrm{mixed}}} = 
\langle\mathsf{p}^2 \rangle_{_{\mathrm{mixed}}} - \langle\mathsf{p} \rangle^{2}_{_{\mathrm{mixed}}},\nn\\
\sigma_{12}|_{_{\mathrm{mixed}}} &=& \tfrac{1}{2}\,\langle(\mathsf{p\,q+q\,p})\rangle_{_{\mathrm{mixed}}} - 
\langle\mathsf{q}\rangle_{_{\mathrm{mixed}}}\,\langle\mathsf{p}\rangle_{_{\mathrm{mixed}}}.
\label{correl-mixed} 
\eea

\hspace{1cm}\begin{table}[]
\centering
\renewcommand{\arraystretch}{1.1}
\begin{tabular}{|C{1.1cm}|C{0.35cm}|C{1.45cm}|C{1.1cm}|C{0.9cm}|C{3.1cm}|C{1.35cm}|C{1.35cm}|C{0.8cm}|C{0.7cm}|C{0.75cm}|}
\hline
$\omega t$ & $2p$ & $ \begin{pmatrix}
\langle \mathsf{q} \rangle \\ \langle \mathsf{p} \rangle
\end{pmatrix} $ &  $\begin{pmatrix}
 \sigma_{11} \\ \sigma_{12} \\ \sigma_{22}
\end{pmatrix}$ & $\widetilde{\vartheta}$ & $\! \! g_k|k=0,1,\ldots,2p-1\!\!\!\!$& $ \begin{pmatrix}
\langle \mathsf{q} \rangle \\ \langle \mathsf{p} \rangle 
\end{pmatrix}_{_{\!\!\!\mathrm{\!\! \! mixed}}}\!\!\!$ &  \hspace{-0.2cm} \!\!\!$\!\!\! \begin{pmatrix}
 \sigma_{11}\\  \sigma_{12} \\ \sigma_{22}
\end{pmatrix}_{_{\!\!\!\mathrm{\!\!\!mixed}\;\;}}\;\;$   & \rotatebox{90}{$S(Q||Q_{_{\mathrm{mixed}}}) \; \; $} 
& \rotatebox{90}{$d_{\mathrm{HS}}(\rho_{\mathcal O}, \rho_{_{\mathrm{mixed}}}^{\hbox{\tiny th}}) \; \;$} 
&\rotatebox{90}{$\dfrac{d_{\mathrm{HS}}(\rho_{\mathcal O}, \rho_{_{\mathrm{mixed}}}^{\hbox{\tiny th}})}{\sqrt{\mathrm{Tr}(\rho_{\mathcal O}(t)^2}} \; \;$}  \\ \hline
     3371      &   4 &     $-0.00062,$ $0 $ &   $ 11.09$, $-0.10$, $9.67$ & $\!\!\! 135.22^{\circ}\;\;\;\;\;\;\;\;\;$ &  1.04081, 1.01542, 1.0404, 1.015   &    $-0.00062$, $0$  &      10.38, $-0.10$, 10.38 & $0.0266$  &  0.518 &  0.732            \\ \hline
286370      &  6     &    0.0083, 0.000012 &  10.36, 0.04, 10.38  & $18^{\circ}$   &  0.9925, 1.02, 0.98139, 0.99, 1.014, 0.99129   &    0.0079, 0.000012  &        
10.31, 0.04, 10.45  & 0.0025   & 0.583 & 0.818          \\ \hline
434450      & 8     &   $ -0.179,$ $-0.033$  &  10.30, $-0.05$, 10.38 &  $26.1^{\circ}$  & 0.986, 0.943, 1, 1, 1, 1.003, 1, 1   &  $-0.174$, $-0.034$                                                                                                        &   10.41, $-0.05$, 10.34 & 0.0049   &   0.604  & 0.832      \\ \hline
\end{tabular}
\caption{Non-Gaussian characteristics of the high entropy states}
\label{H-Entropy}
\end{table}

To emphasize the essential feature of nonclassicality that separates the oscillator state (\ref{o_density_matrix}) at the above high entropy ($S$) configurations with the corresponding statistical mixture of the Gaussian states (\ref{den-th}) we further obtain  the Kullback-Leibler relative entropy [\cite {KL1951}] between  their respective positive semidefinite $Q$-functions. Assuming that the Husimi $Q$-functions $(Q_1(\beta,\beta^*), Q_2(\beta,\beta^*))$ corresponding to their specific quantum density matrices $(\rho_{1}, \rho_{2})$ are known, the nonnegative divergence between the two quasi probability distributions is defined as follows:
\beq
S_{\mathrm{KL}}(Q_1(\beta,\beta^*)||Q_2(\beta,\beta^*)) = \int Q_1(\beta,\beta^*) \log \left(\dfrac{Q_1(\beta,\beta^*)}{Q_2(\beta,\beta^*)}\right) \mathrm{d}^2 \beta.
\label{KL_Q1Q2}
\eeq
The construction of the divergence between the two states may be thought of as the quasi classical analog of the quantum relative entropy
$S(\rho_{1}||\rho_{2}) = \mathrm{Tr} [\rho_{1} (\log \rho_{1} - \log \rho_{2})]$ [\cite{NC2000}] much in the sense that Wehrl entropy
$S_W$ is regarded [\cite{KMOS2009}] a qualitatively similar approximation to the von Neumann entropy $S$. We make the identification of the oscillator 
$Q$-function (\ref{Q_factorized}) with $Q_1(\beta,\beta^*)$ and that of the statistical mixed state (\ref{Q_th_mix}) with 
$Q_2(\beta,\beta^*)$.

\par

The oscillator density matrix (\ref{o_density_matrix}) corresponding to $2 p$ squeezed kitten states with near-maximal entropy configuration arrived at times 
$T_{\hbox{\tiny long}}/p$ (up to a period $T_{\hbox{\tiny long}}$) is now compared with the fiducial density 
matrix (\ref{den-th}). Employing the angular distribution function $\mathcal{P}(\theta)$ for the respective cases, 
the initial rotation angle $\widetilde{\vartheta}$ for the statistical mixture in (\ref{den-th}) is selected so that the overlap between two configurations is maximized. The angular function $\mathcal{P}(\theta)$ for the statistical mixture (\ref{den-th}) is plotted in Figs. \ref{SW_min} 
({\sf d, h, l}) via dotted green lines. The analytic expression may be easily recovered from (\ref{den-th}). 
We do not explicitly quote it here. The  
coefficients $\{g_{k}| k = 0, 1,\ldots, 2p -1\}$ in the density matrix (\ref{den-th}) are now \textit{independently} 
varied so that the first two quadrature moments and the covariance matrix resulting therefrom are nearly equal.
Therefore, the semiclassical features of two density matrices (\ref{o_density_matrix}, \ref{den-th}) are almost 
identical. The nearly indistinguishable quadrature moments 
up to the second order 
for the two density matrices in comparison confirm this (Table \ref{H-Entropy}). It is further 
corroborated by the approximately null value of the Kullback-Leibler divergence (\ref{KL_Q1Q2}) between the two cases. 
However, the distinction between the two examples appear due to the inherent non-Gaussian nature of the oscillator
state (\ref{o_density_matrix}). The statistical mixture (\ref{den-th}) of the displaced squeezed thermal density
matrices has completely positive $W$-distribution (\ref{W_th_mix}), whereas the 
oscillator density matrix 
(\ref{o_density_matrix}) considered, in particular,  at the times corresponding to near maximal entropy 
($\omega t = 3371, \omega t = 286370, \omega t = 434450$ in Fig. \ref{kittenDouble} ({\sf a}), ({\sf b}), ({\sf c}),
respectively) display large negativity $\delta_{W}$  that points towards its highly nonclassical nature of the probed 
states. The essential non-Gaussianity of the density matrix (\ref{o_density_matrix}) in the said maximal entropy region is,
however,  well-accounted by the Hilbert-Schmidt distance 
$d_{\mathrm{HS}}(\rho_{\mathcal O}, \rho_{_{\mathrm{mixed}}}^{\hbox{\tiny th}})$ at the relevant times. 
The Table \ref{H-Entropy} lists a summary of our 
study of nonclassicality of the high entropy states considered above. In these cases we note that subject to the 
equality of the Gaussian properties of the density matrices (\ref{o_density_matrix}) and (\ref{den-th}), we enumerate 
the least Hilbert-Schmidt distance between them. The Table \ref{H-Entropy} reveals that despite the close kinship in the
Gaussian properties, the relatively high magnitudes of the dimensionless ratio
$d_{\mathrm{HS}}(\rho_{\mathcal O}, \rho_{_{\mathrm{mixed}}}^{\hbox{\tiny th}})/\sqrt{\mathrm{Tr}(\rho_{\mathcal O}(t)^2)} \sim 1$ 
point towards prominent nonclassicality of the oscillator density matrix, even when it is far away from a pure state.
In this sense the Hilbert-Schmidt distance $d_{\mathrm{HS}}(\rho_{\mathcal O}, \rho_{_{\mathrm{mixed}}}^{\hbox{\tiny th}})$ provides the measure of the quantum properties of the oscillator state (\ref{o_density_matrix}). We also note that, when  compared (Table \ref{H-Entropy}) with the oscillator density matrix (\ref{o_density_matrix}) the reference density matrix 
$\rho_{_{\mathrm{mixed}}}^{\hbox{\tiny th}}$ (\ref{den-th}) yields nearly equal coefficients only for the null temperature: $\bar{n} = 0$. This is expected as we do not consider the finite temperature effects in the time evolution of the qubit-oscillator system in the present analysis.

\section{Conclusion}
\label{conclude}

Employing the adiabatic approximation we have studied a qubit-oscillator bipartite system in the presence of a static bias term for the strong coupling regime. 
Starting with a hybrid squeezed cat type of state we obtain the evolution of the qubit and the oscillator reduced density matrices. The oscillator density matrix
furnishes the diagonal $P$-representation on the phase space. The rapidly oscillating derivatives of the $\delta$-functions present  in the $P$-representation
make it highly singular. Two successive smoothing of the singular $P$-representation via Gaussian kernels generate first the Wigner $W$-distribution, and subsequently 
the nonnegative Husimi $Q$-function. The interference between the quantum fluctuations cause the quasi probability distribution $W$ to develop negative values.
Its negativity measure $\delta_{W}$ marks a departure of the state from 
Gaussian configurations. The nonnegative $Q$-function yields the Wehrl entropy $S_{Q}$ that measures the delocalization in the oscillator phase space. The qubit-oscillator interaction establishes the presence of multiple time scales that triggers some novel features. In the strong coupling regime $\lambda \lessapprox 0.1 \,\omega$ the fluctuations 
with the frequencies $O(x^{2} \widetilde{\Delta})$ institute a long term quasi periodicity  in the system with the time period given by
$T_{\hbox{\tiny long}} \sim O((x^{2} \widetilde{\Delta})^{-1})$. Superimposed on them are the short time quantum fluctuations with 
period $T_{\hbox{\tiny short}} \sim O((x \widetilde{\Delta})^{-1})$ that effect an energy transfer between the qubit and the oscillator. At the rational submultiples of 
the long time period, say  $T_{\hbox{\tiny long}}/p$, we observe that oscillator states endowed with the local minima of entropies $(S, S_{Q})$ develop during the course 
of $T_{\hbox{\tiny short}}$ oscillations. These almost pure states reside in the Hilbert space in the close neighborhood  of  $p$ superposed squeezed kitten states that are equi-rotated in the phase space. This is
established  by considering the numerical variation of the Hilbert-Schmidt distance over an ensemble of states. The oscillations with period $T_{\hbox{\tiny short}}$ now 
cause a bifurcation of the peaks in the phase space so that at the instances characterized by the near maximal values of the entropies $(S, S_{Q})$  a phase space 
distribution of $2 p$ equally rotated distinguished squeezed 
kitten states develop. We compare these states with the statistical mixture of Gaussian squeezed states, chosen on the basis of near equality of the first two 
quadrature moments with the former. As a further check the Kullback-Leibler divergence based on the smoothed nonnegative $Q$-functions for the two states in comparison 
is found to have almost null value. The non-Gaussianity of the oscillator state becomes manifest  as its Hilbert-Schmidt distance with the Gaussian reference state becomes significantly large. The bifurcation and 
rejoining of the squeezed kitten states may be of practical significance in building quantum computational network. The qubit-oscillator
bipartite system with time-dependent coupling may be useful in this context.

\section*{Acknowledgements}
One of us (MB) acknowledges the support from the UGC (India) under non-NET fellowship, and a
URF from the University of Madras. Another author (BVJ) is  supported by the UGC (India) under the Maulana Azad National Fellowship  scheme.

\end{document}